\documentclass[aps,floatfix,nofootinbib,notitlepage]{revtex4-1}
\usepackage[utf8]{inputenc}

\usepackage{hyperref}
\usepackage{graphicx}

\makeatletter
\let\MYcaption\@makecaption
\makeatother
\usepackage{subcaption}
\makeatletter
\let\@makecaption\MYcaption
\makeatother

\usepackage{bm}
\usepackage{amsmath}
\usepackage{amssymb}

\newcommand{\beq}{\begin{eqnarray}}
\newcommand{\eeq}{\end{eqnarray}}

\newcommand{\tr}{{\rm tr}  }
\newcommand{\nn}{\nonumber \\}

\begin{document}

\title{Non-global logarithms in hadron collisions at \texorpdfstring{$N_c=3$}{Nc=3}}

\author{Yoshitaka Hatta}
\email{yhatta@bnl.gov}
\affiliation{Physics Department, Brookhaven National Laboratory, Upton, NY 11973, USA}
\affiliation{RIKEN BNL Research Center, Brookhaven National Laboratory, Upton, NY 11973, USA}

\author{Takahiro Ueda}
\email{tueda@st.seikei.ac.jp}
\affiliation{Faculty of Science and Technology, Seikei University, Musashino, Tokyo 180-8633, Japan}

\begin{abstract}
We calculate the rapidity gap survival probability associated with the Higgs decay and Higgs plus dijet production  in proton-proton collisions by resumming the leading non-global logarithms
without any approximation to the number of colors. For dijet production, depending on partonic subprocesses, the probability involves various  `color multipoles', i.e., the product of 4 ($qq\to qq)$ or  6 ($qg\to qg$) or 8 ($gg\to gg$) Wilson lines.
We calculate all these multipoles for a fixed dijet configuration and discuss the factorization of higher multipoles into lower multipoles as well  as the validity of the large-$N_c$ approximation. 
\end{abstract}

\maketitle

\section{Introduction}
\label{sec:intro}

Recently, there have been a lot of activities in developing  Monte Carlo algorithms for simulating  parton showers  beyond the large-$N_c$ (leading-$N_c$) approximation where $N_c=3$ is the number of colors \cite{Platzer:2012np,Nagy:2015hwa,Dasgupta:2018nvj,Forshaw:2019ver,Nagy:2019rwb,Hoeche:2020nsx,Dasgupta:2020fwr,Balsiger:2020ogy,DeAngelis:2020rvq,1831912}. Traditionally, in most event generators, the large-$N_c$ approximation has been the only practical way to keep track of the color indices of many partons involved    \cite{Hoche:2014rga}. Any attempt to include $N_c$-suppressed corrections will be met with fierce computational challenges which might require  a drastic overhaul of the existing  approaches. Yet, such efforts seem to be unavoidable in view of the   ever-increasing demand for precision  at the LHC and future collider experiments.

Among other observables,  the finite-$N_c$ corrections are particularly important but difficult to quantify for the so-called  non-global observables \cite{Dasgupta:2001sh,Dasgupta:2002bw} which are sensitive to the wide-angle emission of soft gluons in restricted regions of phase space. The resummation of  non-global logarithms has been originally done in the large-$N_c$ approximation \cite{Dasgupta:2001sh,Dasgupta:2002bw,Banfi:2002hw} where it has been observed  that including the finite-$N_c$ corrections is highly nontrivial even at the leading-logarithmic level.  Therefore, an accurate description of non-global observables serves as an important litmus test  for any event generator purported to contain `full-color' parton showers. 

In order to carry out such a test, it is  necessary to provide  benchmark finite-$N_c$ results that one can compare to. In \cite{Hatta:2013iba}, we have developed a framework to resum non-global logarithms at $N_c=3$ by improving and completing the earlier attempt   \cite{Weigert:2003mm}. It is based on an analogy (actually, equivalence \cite{Hatta:2008st,Caron-Huot:2015bja,Neill:2020bwv}) to the resummation of logarithms $(\ln 1/x)^n$ in small-$x$ QCD, and is formulated as the random walk of Wilson lines in the color SU(3)  space \cite{Blaizot:2002np}.   Numerical results are so far available only for two observables in $e^+e^-$ annihilation:  interjet energy flow  \cite{Hatta:2013iba} and the hemisphere jet mass distribution \cite{Hagiwara:2015bia}. The impact of the finite-$N_c$ corrections has been found to be somewhat larger for the latter observable, but overall, deviations from the large-$N_c$ results are not spectacular  at least in   the phenomenologically relevant region of parameters. Yet, it has been already envisaged in \cite{Hatta:2013iba} that the finite-$N_c$ effects will be stronger in hadron-hadron collisions. It is the purpose the present paper to demonstrate that our approach can be practically applied to hadron collisions where it is probably  most useful.
We do so by explicitly computing two observables relevant to the $pp$ collisions at the LHC\@. These are the rapidity gap survival  (or `veto')  probabilities in  the  Higgs boson decay $H\to gg$ (Section~\ref{sec:higgs-decay}) and in Higgs plus dijet production $pp \to HjjX$ (Section~\ref{sec:higgs-dijet}). The relevant logarithms are of the form  $(\alpha_s \ln Q/E_{out})^n$ where $Q$ is the hard scale (Higgs mass or jet transverse momentum) and $E_{out}\ll Q$ is the veto scale. For  previous related studies in $pp\to jjX$ in the large-$N_c$ approximation, see \cite{Hatta:2013qj,Hatta:2009nd}.

A novel feature of hadron-hadron collisions  as opposed to $e^+e^-$ annihilation is that the gap survival probability is given by `color multipoles'---the correlation functions of up to eight Wilson lines in the fundamental representation 
\beq
{\rm tr}(U_\alpha U_\beta^\dagger U_\gamma U_\delta^\dagger), 
\quad {\rm tr}(U_\alpha U_\beta^\dagger U_\gamma U^\dagger_\delta U_\epsilon U_\zeta^\dagger), \quad 
{\rm tr}(U_\alpha U_\beta^\dagger){\rm tr}( U_\gamma U^\dagger_\delta U_\epsilon U_\zeta^\dagger U_\eta U^\dagger_\theta),\quad \cdots \label{multi}
\eeq
where each Wilson line $U_\alpha$   is associated with   a hard parton moving in direction $\alpha$ involved in $2\to 2$ subprocesses.  The maximal number is eight  because  a gluon counts as two Wilson lines. In $e^+e^-$ annihilation, one only has to deal with the color dipole ${\rm tr}(U_\alpha U_\beta^\dagger)$ corresponding to the $q\bar{q}$ pair in the final state. While the calculation of higher multipoles is  more cumbersome, it  does not pose any particular problems. We shall  present the first results on 
the resummation of non-global logarithms for these multipoles valid exclusively at $N_c=3$.  

It is worthwhile to mention that our calculation can be viewed as the timelike counterpart of the corresponding spacelike problem, namely,   the resummation of small-$x$ (`BFKL') logarithms for the color quadrupole ${\rm tr}(UU^\dagger UU^\dagger)$ and other higher multipoles relevant to high energy scattering  \cite{Balitsky:1995ub,Kovchegov:2008mk,Dominguez:2011gc,Dumitru:2011vk,Marquet:2016cgx}. In that context, $U_\alpha$ is a  Wilson line along the light-cone representing the final state interaction. The label $\alpha$  denotes a point on the transverse plane mapped from the sphere in the timelike problem  via the stereographic projection \cite{Hatta:2008st}.    We shall contrast our results with those in the small-$x$ literature when we discuss the  factorization of higher multipoles into lower multipoles as well as the validity of the mean field  approximation.

\section{Resummation strategy}
\label{sec:strategy}

In this section we briefly recapitulate the  procedure for resumming non-global logarithms at $N_c=3$ pioneered  in \cite{Weigert:2003mm} and completed in  \cite{Hatta:2013iba,Hagiwara:2015bia}. We first divide the $4\pi$ solid angle $(\cos \theta,\phi)$ into the `in'-region which contains hard partons (incoming partons and outgoing jets) and the `out'-region where measurements are done. We fix the out-region to be the mid-rapidity region defined by 
\beq
\cos \theta_{in} > \cos\theta > -\cos\theta_{in}, \qquad 2\pi >\phi>0.
\label{out}
\eeq
The in-region is the  complement of this. The next step is to discretize the in- and out-regions,  and on each grid point $\alpha$ in the in-region, we put an SU(3) matrix.
\beq
U(\cos \theta_\alpha, \phi_\alpha,\tau) =U_\alpha(\tau).
\eeq
 Each matrix evolves in `time' 
\beq
\tau=\begin{cases} \frac{\alpha_s}{\pi}\ln \frac{Q}{E_{out}} \qquad ({\rm fixed\ coupling}) \\
\frac{6}{11N_c-2n_f}\ln \frac{\ln Q/\Lambda_{QCD}}{\ln E_{out}/\Lambda_{QCD}} \qquad ({\rm running\  coupling})
\end{cases}
\label{fix}
\eeq
where $n_f$ is the number of flavors, $Q$ is the typical hard scale (like the transverse momentum of jets in the in-region) and $E_{out}\ll Q$ is the maximum total energy emitted into the out-region. 
The initial condition is $U_\alpha=1$ for all $\alpha$. In each step of evolution, $U_\alpha$ changes as 
\beq
U_\alpha(\tau + \epsilon)=e^{iS_\alpha^{(2)}}e^{iA_\alpha}U_\alpha(\tau)e^{iB_\alpha}e^{iS^{(1)}_\alpha} \label{chan}
\eeq
where 
\beq
S_\alpha^{(i)}&=&\sqrt{\frac{\epsilon}{4\pi}} \int_{out} d\Omega_\gamma \frac{({\bf n}_\alpha -{\bf n}_\gamma)^k}{1-{\bf n}_\alpha \cdot {\bf n}_\gamma}t^a \xi^{(i)k}_{\gamma a} \qquad (i=1,2) \nn 
A_\alpha &=& -\sqrt{\frac{\epsilon}{4\pi}} \int_{in} d\Omega_\gamma \frac{({\bf n}_\alpha -{\bf n}_\gamma)^k}{1-{\bf n}_\alpha \cdot {\bf n}_\gamma} U_\gamma t^a U^\dagger_\gamma \xi_{\gamma a}^{(1)k}
\label{a} \nn 
B_\alpha &=&\sqrt{\frac{\epsilon}{4\pi}} \int_{in} d\Omega_\gamma \frac{({\bf n}_\alpha -{\bf n}_\gamma)^k}{1-{\bf n}_\alpha \cdot {\bf n}_\gamma} t^a \xi^{(1)k}_{\gamma a}
\eeq
 ${\bf n}_\alpha$ is the three-dimensional unit vector in  direction $\alpha$. The solid angle integrals are restricted to the in- or out-region as indicated. $t^{a=1,2,..,8}$ are the SU(3) generators normalized as ${\rm tr}t^a t^b=\delta^{ab}/2$.   $\xi^{(1,2)}$ are Gaussian white noises randomly generated at every time step and at every grid point (not just in the in-region where $U$'s are defined). They are characterized by the correlator
\beq
\langle \xi_{\gamma a}^{(i)k}(\tau) \xi^{(j)l}_{\gamma'b}(\tau')\rangle 
= \delta^{ij}\delta_{\tau,\tau'}\delta(\Omega_\gamma-\Omega_{\gamma'})\delta_{ab}\delta^{kl}
\eeq
where $\langle ...\rangle$ denotes averaging over events. 

Physically, $U$'s are Wilson lines from the origin to spatial infinity,   representing the primary hard partons as well as the secondary gluons that are emitted in the in-region.  Non-global logarithms arise from the region of phase space where the successive emissions  are strongly ordered in energy. In each emission, the parent parton can be treated as a Wilson line in the spirit of the eikonal approximation. Note that gluons should be described by Wilson lines in the adjoint representation $\tilde{U}_\alpha$, but they can always be reduced to those in the fundamental representation via the identity
\beq
\tilde{U}_\alpha^{ab}t^b = U^\dagger_\alpha t^a U_\alpha .
\eeq
When a soft gluon is emitted, each Wilson line receives random kicks in the color space as indicated by the various factors in (\ref{chan}). Roughly speaking, $S_\alpha$ generates the Sudakov logarithms, and $A_\alpha$ and $B_\alpha$ accounts for the non-global logarithms, although this distinction cannot be made clear-cut.

It is important to mention that  one should really think of $U_\alpha$ as the product 
\beq
U_\alpha \to  V_\alpha^\dagger U_\alpha  \label{pro}
\eeq
where $V_\alpha$ is the same Wilson line as $U_\alpha$, but defined in the complex-conjugate amplitude. Namely, we are considering the evolution of  probabilities rather than  amplitudes, see \cite{Weigert:2003mm} for a careful discussion on this point. In the following, we keep using the simpler notation $U_\alpha$, but what we  actually mean is the product (\ref{pro}).

A peculiar feature of the evolution (\ref{chan}) is that, even though we directly deal with probabilities, the actual evolution looks like being  implemented at the amplitude level as can be seen by noticing  that the integration kernel of $S_\alpha$, $A_{\alpha}$ and $B_\alpha$ in (\ref{a}) is the `square-root' of the soft emission probability\footnote{Note that the kernel is different from the `naive' square root $\frac{p_\alpha^\mu}{p_\alpha \cdot k_\gamma}$ which is the usual eikonal factor. This does not work in the present scheme as explained in  \cite{Hatta:2013iba}. }
\beq
\frac{p_\alpha \cdot p_\beta}{p_\alpha\cdot k_\gamma p_\beta\cdot k_\gamma} \label{we}
\eeq
At the end of the evolution, these kernels are `glued together' by  averaging over noises to form the probability (\ref{we}).

With this setup, a typical simulation goes as follows.  We evolve $U$'s in $\tau$ for many different realizations of random noises (`trajectories') up to a desired time $\tau$. Phenomenologically, $\tau \sim 0.5$ at most.  We then calculate color multipoles  such as
\beq
{\rm tr}(U_\alpha U_\beta^\dagger),\qquad  {\rm tr}(U_\alpha U_\beta^\dagger U_\gamma U^\dagger_\delta), \quad \cdots
\eeq
in each trajectory and average them over many  (practically more than $500$) trajectories. The results are related to the `veto' cross section, namely, the probability  that the total energy emitted from color-singlet antennas $\alpha\beta$, $\alpha\beta\gamma\delta$,..  into the out-region is less  than $E_{out}$ \cite{Weigert:2003mm,Hatta:2013iba,Caron-Huot:2015bja}. Both the leading Sudakov and non-global logarithms  $(\alpha_s \ln Q/E_{out})^n$ are included to all orders, and no approximation is made as to the number of colors  $N_c=3$.

\section{Higgs decaying into two gluons}
\label{sec:higgs-decay}

So far, all-order, finite-$N_c$ results are available only for two specific observables in $e^+e^-$ annihilation: Interjet energy flow \cite{Hatta:2013iba}   and the  hemisphere jet mass distribution     \cite{Hagiwara:2015bia}. In this and the next section, we shall enlarge this list by including two hard processes relevant to $pp$ collisions at the LHC\@. First,  we consider the jet veto cross section associated with the decay of the Higgs boson $H\to gg$ where the Higgs is created by the weak interaction  so that there is no QCD radiation from the initial state. This process has been recently studied in  \cite{DeAngelis:2020rvq} as a test case to resum non-global logarithms including finite-$N_c$ corrections in a different framework. Following this reference, we work in  the Higgs rest frame and  the back-to-back gluons are moving in directions $\theta=0,\pi$. We then define the in- and out-regions as in (\ref{out}). 
The gap survival probability is given by  
\beq
P_H(\tau)=\frac{1}{N_c^2-1}\langle {\rm tr} (\tilde{U}_0(\tau) \tilde{U}_\pi^\dagger(\tau))\rangle_\xi, %
\label{tau}
\eeq
where  $\tau$ is given by (\ref{fix}) with $Q=M_H$, the Higgs boson mass.  $\tilde{U}$ is an $8\times 8$ matrix in the adjoint representation of SU(3)  appropriate for the outgoing gluons. Note that the same formula can be used to compute the veto cross section associated with the production $gg\to H$ with a subsequent non-hadronic decay of the Higgs boson. However, in this case there are  extra complications from the so-called `super-leading' logarithms \cite{Forshaw:2006fk,Forshaw:2008cq} which arise when the final state gluons become collinear to the incoming gluons. They cannot be resummed in the present framework  because we neglect the `$i\pi$-terms' in the soft functions \cite{Forshaw:2006fk,Forshaw:2008cq,Nagy:2019rwb}. Therefore, while the result below is relevant to both processes $H\to gg$ and $gg\to H$, care must be taken when applying it to the latter.

To evaluate $P_H(\tau)$, we use the fact  that any SU(3) matrix in the adjoint representation $\tilde{U}$ can be identically written in terms of the corresponding matrix in the fundamental representation $U$ as
\beq
\tilde{U}^{ab} = 2{\rm tr}(U^\dagger t^a U t^b), \qquad  {\rm tr}\, \tilde{U} = |{\rm tr}\, U|^2-1\,,
\eeq
so that
\beq
P_H(\tau)&=&\frac{1}{N_c^2-1}
\left\langle |{\rm tr} \, U_0  U_\pi^\dagger|^2-1 \right\rangle_\xi \nonumber \\  &=& \frac{1}{N_c^2-1}\Bigl\langle   (\mathfrak{Re}\, {\rm tr} \, U_0 U_\pi^\dagger)^2 +  (\mathfrak{Im}\, {\rm tr} \, U_0 U_\pi^\dagger)^2  -1\Bigr\rangle_\xi\,. \label{higg}
\eeq
The problem has thus reduced to calculating the dispersion of the real and imaginary parts of color dipoles.
 As observed in \cite{Hatta:2013iba}, the imaginary part vanishes (within errors) after averaging over noises $\langle \mathfrak{Im}\,{\rm tr}U_0U_\pi^\dagger\rangle_\xi =0$.  However, there are huge event-by-event fluctuations which lead to the nonvanishing dispersion $\langle (\mathfrak{Im}\,{\rm tr}U_0U_\pi^\dagger)^2\rangle_\xi \neq 0$. This   actually plays a crucial role in the present calculation.

\begin{figure}
  \includegraphics[width=0.8\linewidth]{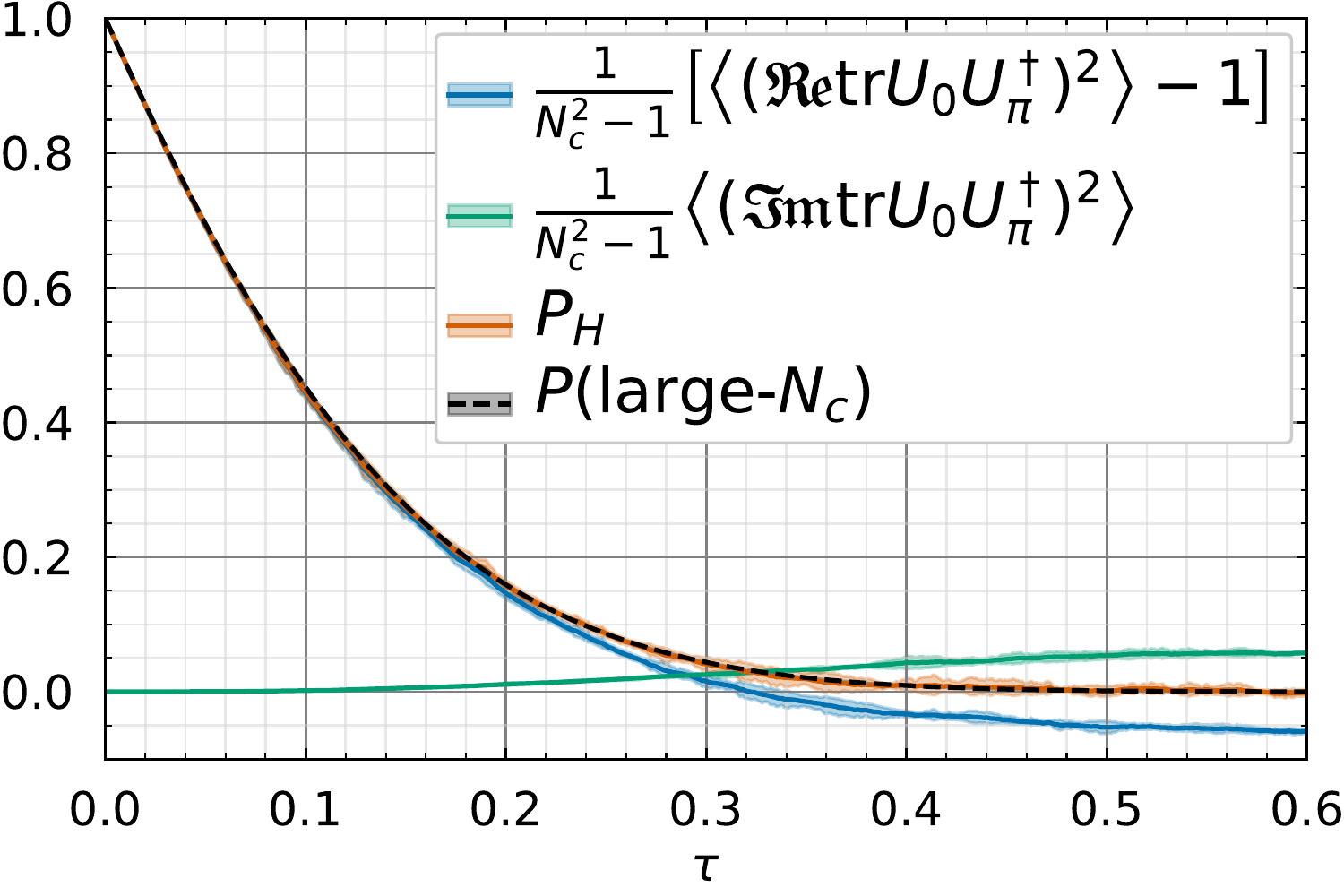}
\caption{Gap survival probability  (\ref{higg}) in $H\to gg$ with $\theta_{in}=\pi/3$  as a function of $\tau$. The large-$N_c$ result is constructed from the solution of (\ref{bms}) with $C_F=N_c/2$.
}
\label{fig}
\end{figure}

\begin{figure}
  \centering
  \begin{subfigure}{0.49\linewidth}
    \centering
    \includegraphics[width=\textwidth]{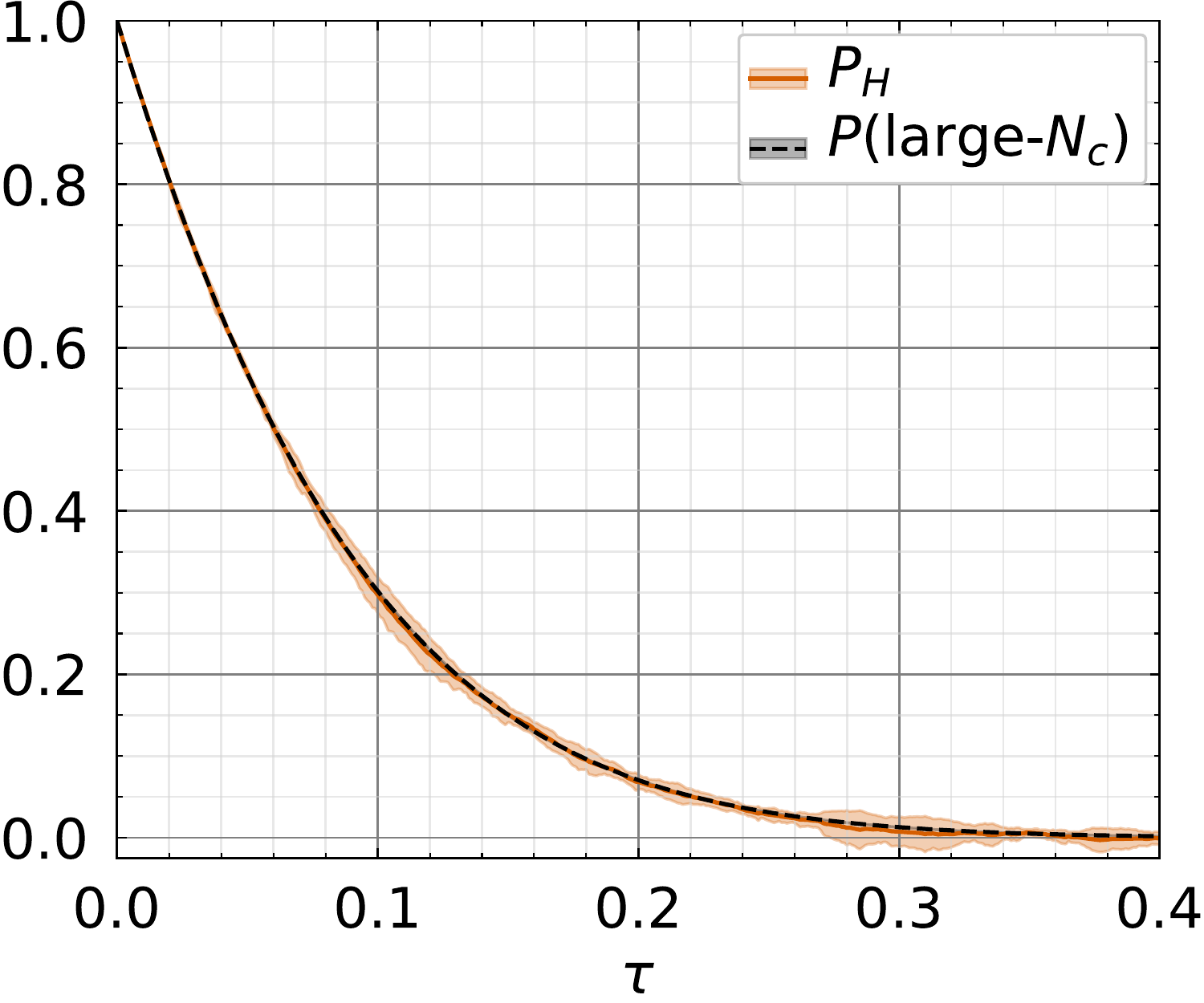}
    \caption{}
    \label{fig:2a}
  \end{subfigure}
  \hfill
  \begin{subfigure}{0.49\linewidth}
    \centering
    \includegraphics[width=\linewidth]{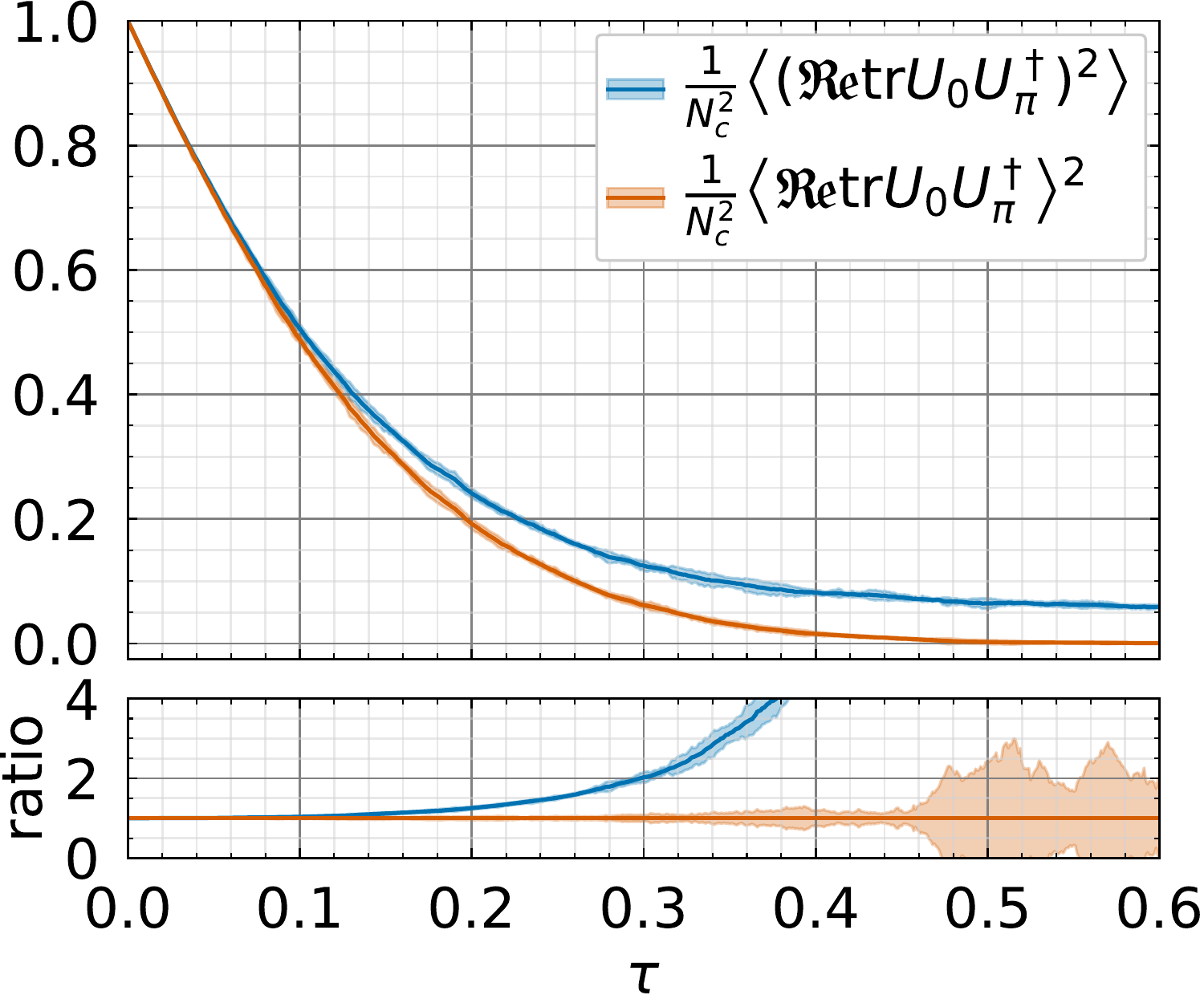}
    \caption{}
    \label{fig:2b}
  \end{subfigure}
  \caption{(a) Same as Fig.~\ref{fig} except that here $\theta_{in}=\pi/4$; (b) Plots of the two terms in (\ref{meanfield}), both normalized to unity at $\tau=0$.}
\end{figure}

 We use a uniform $80\times 60$ lattice in the $(\cos \theta,\phi)$ plane and set $\theta_{in}=\pi/3$.  The time step is chosen to be $\epsilon=5\times 10^{-5}$, and we perform the `reunitarization' of all the $U$'s  after every 100 steps of iteration.   The result, averaged over 3000 trajectories, is shown in Fig.~\ref{fig}. Each error band represents the sum of statistical  and
systematic errors.
The latter are estimated by performing  simulations with $\epsilon=10^{-4}$ and also on a $60\times 40$ lattice with $\epsilon=5\times 10^{-5}$ (all 3000 trajectories). 
In Fig.~\ref{fig:2a}, we show the result for a different opening angle  $\theta_{in}=\pi/4$ in order to facilitate  comparison with Ref.~\cite{DeAngelis:2020rvq}. 

From Fig.~\ref{fig}, we see that, without the contribution from the imaginary part $\sim(\mathfrak{Im}\,{\rm tr} UU^\dagger)^2$, $P_H(\tau)$ becomes negative. At large-$\tau$, $P_H(\tau)$ goes to zero due to an almost exact cancellation between  the real and imaginary contributions. 
To further appreciate the importance of the dispersion, in Fig.~\ref{fig:2b}, we compare \beq
\left\langle \left(\frac{1}{N_c}\mathfrak{Re}\, {\rm tr} U_0U_\pi^\dagger \right)^2\right\rangle \qquad {\rm vs.} \qquad  
\left\langle \frac{1}{N_c}\mathfrak{Re}\, {\rm tr} U_0U_\pi^\dagger \right\rangle^2. \label{meanfield}
\eeq
In the usual large-$N_c$ argument, the two quantities are approximately equal up to corrections of order $1/N_c^2\sim 10$\%. However, this is clearly not the case except in the small-$\tau$ region. Already around $\tau\sim 0.3$, the corrections reach 100\%, and the ratio blows up as $\tau$ gets larger. 
It is tempting to explain this by saying that  the probability $P_H$ itself becomes of order $1/N_c^2\sim 0.1$ in this region, so the finite-$N_c$ corrections become an ${\cal O}(1)$ effect. However, our interpretation is different. 
The enhancement shown in  Fig.~\ref{fig:2b} is   reminiscent of that of the dipole pair distribution in Mueller's dipole model \cite{Mueller:1994jq}, both for the spacelike \cite{Hatta:2007fg,Avsar:2008ph} and timelike  \cite{Avsar:2009yb} parton showers. As demonstrated  in these references,  drastic violations of the `mean field approximation' $\langle AB\rangle\approx \langle A\rangle \langle B\rangle$  can result from the spatial correlation among soft gluons induced by the small-$x$ evolution, and this has nothing to do with the number of colors. 
To support this  interpretation, in the next section we show that the quality of the   approximation $\langle AB\rangle\approx \langle A\rangle \langle B\rangle$ crucially depends on the spatial configuration of dipoles. \\

Finally, the black dashed curves in Fig.~\ref{fig} and Fig.~\ref{fig:2a}  are the square of the large-$N_c$ result by Dasgupta and Salam  \cite{Dasgupta:2002bw}, or equivalently
 the solution of the Banfi-Marchesini-Smye (BMS) equation \cite{Banfi:2002hw}
\beq
\frac{\partial}{\partial \tau}P_{\alpha\beta}(\tau)&=&-2C_F\int_{out} \frac{d\Omega_\gamma}{4\pi}\frac{1-\cos\theta_{\alpha\beta}}{(1-\cos\theta_{\alpha\gamma})(1-\cos\theta_{\gamma\beta})}P_{\alpha\beta}\nn
&& \qquad + N_c\int_{in}\frac{d\Omega_\gamma}{4\pi} \frac{1-\cos\theta_{\alpha\beta}}{(1-\cos\theta_{\alpha\gamma})(1-\cos\theta_{\gamma\beta})} (P_{\alpha\gamma}P_{\gamma\beta}-P_{\alpha\beta}) \label{bms}
\eeq
with $C_F=\frac{N_c^2-1}{2N_c}\approx N_c/2$. $P_{\alpha\beta}$ is the gap survival probability for a quark dipole corresponding to our  $\langle \frac{1}{N_c}{\rm tr}(U_\alpha U_\beta^\dagger)\rangle$. In Fig.~\ref{fig} and Fig.~\ref{fig:2a}, we have plotted\footnote{To solve (\ref{bms}), we perform the integral in the out-region (Sudakov term)  analytically. The in-region integral is done on a 160$\times$120 lattice in $(\cos \theta,\phi)$.  Systematic errors are estimated from solutions on coarser lattices.}
\beq
P(\text{large-}N_c)\equiv (P_{0\pi})^2, \label{sud}
\eeq
with $C_F=\frac{3}{2}$. 
Somewhat surprisingly, we find an almost perfect agreement $P_H\approx P({\rm large}N_c)$.\footnote{We are indebted to Gavin Salam for making this observation and allowing us to show it in this paper. Essentially the same result has been obtained in his parton shower  framework \cite{1831912} which correctly includes full-color results to ${\cal O}(\alpha_s^2)$.}  A possible explanation may be as follows. The probability $P_H$ consists of the Sudakov  and non-global parts. The Sudakov part is just the exponential of the one-loop contribution which is  proportional to $C_A=N_c$ for a gluon dipole and $C_F\approx N_c/2$ for a quark dipole. Thus, the relation $P_H=P(\text{large-}N_c)$ holds exactly for the Sudakov part. The non-global part starts at two-loops ${\cal O}(\alpha_s^2)$, and is propotional to $C_A^2=N_c^2$ for a gluon dipole and $C_AC_F\approx N_c^2/2$ for a quark dipole. If one assumes that this leading term exponentiates (which is nontrivial), and the higher-order terms are  not important or follow a similar pattern (also nontrivial), the relation $P_H\approx P(\text{large-}N_c)$ holds also for the non-global part. It is highly nontrivial to explain this relation in our approach which only deals with matrices in the fundamental representation. In particular, the large violation of the mean field approximation Fig.~\ref{fig:2b} is essential to achieve $P_H\approx P(\text{large-}N_c)$. We shall encounter even more nontrivial  relations to the large-$N_c$ result in the next section.

\section{Jet veto in Higgs plus dijet production}
\label{sec:higgs-dijet}

We now turn our attention to the more  interesting but difficult problem of hadron collisions with $2\to 2$ hard parton subprocesses. In this case, there are four primary partons (quarks or gluons) in the initial and final states, and the emission of soft gluons  from this four-pronged antenna is obviously much more complicated than the previous examples. Nevertheless, the resummation of the Sudakov logarithms can be done (at finite-$N_c$) using the techniques of the soft anomalous dimension  \cite{Kidonakis:1998nf}.  The non-global logarithms are parametrically of the same order, but their resummation has  been done only in the large-$N_c$ approximation for dijet production at the LHC  \cite{Hatta:2013qj}. 
In this section, we perform, for the first time, the leading-logarithmic resummation of non-global logarithms for $2\to 2$ scatterings at finite-$N_c$, taking  Higgs plus dijet production in $pp$ collisions at the LHC as a concrete example. We however have to sacrifice the super-leading logarithms  which are relevant to the present problem since there are hard partons in both the initial and final states. We leave this to future work.

 \begin{figure}
  \includegraphics[width=1\linewidth]{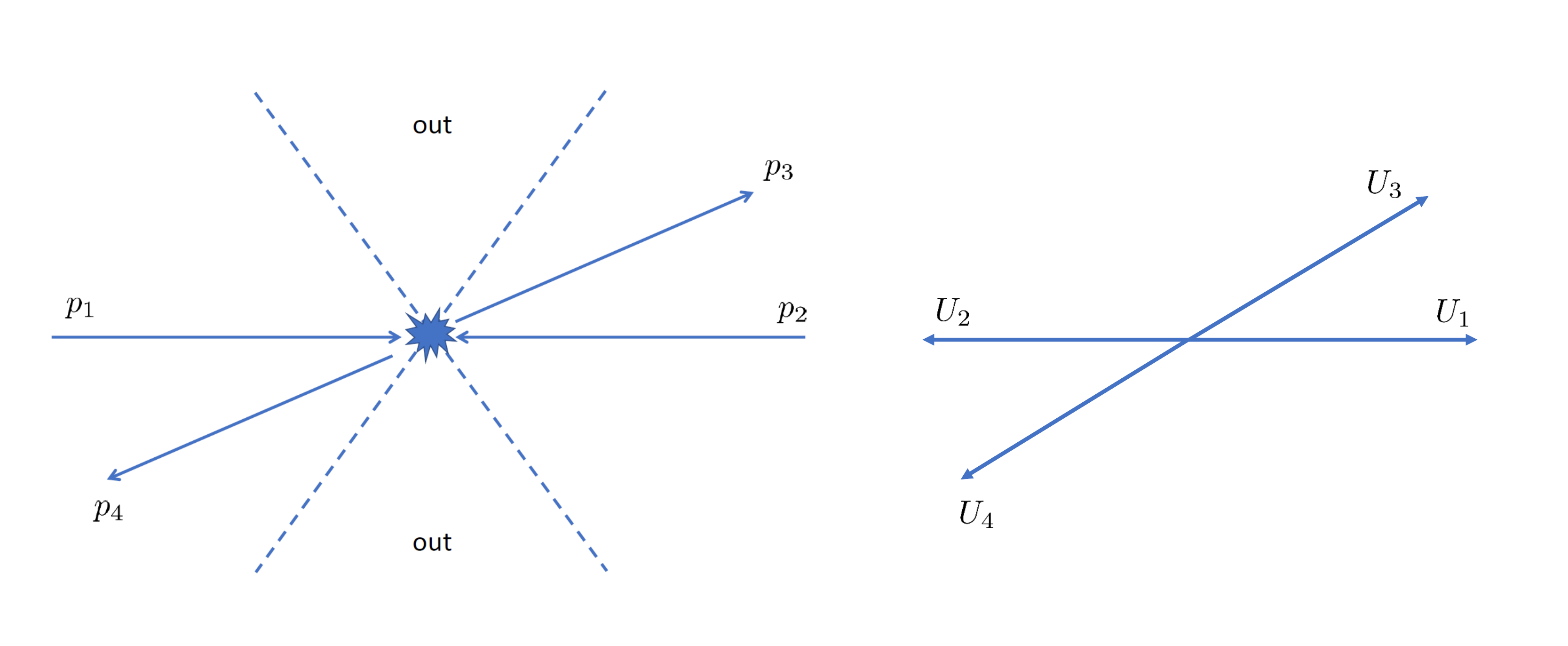}
\caption{ $2\to 2$ parton scattering with momentum assignments (left) and the  corresponding antenna structure (right).
}
\label{2to2}
\end{figure}

\subsection{\texorpdfstring{$qq\to qqH$}{q q -> q q H} %
}

Let us first consider the simplest channel  $q_i(p_1)q_j(p_2) \to q_k(p_3)q_l(p_4) H$ where $i,j,k,l=1,2,3$ are color indices.   The outgoing quarks (or antiquarks) with momenta $p_3$, $p_4$ are back-to-back and detected as two jets in the forward and  backward directions, see Fig.~\ref{2to2}. 
The radiation pattern is  sensitive to how  color flows in the $2\to 2$ scattering. 
Following  \cite{Forshaw:2007vb}, we use the eikonal approximation  $\bar{u}(p_3)\gamma^\mu u(p_1) \approx 2p^\mu_1$ and parameterize the leading-order amplitude as 
\beq
{\mathcal M}_{ijkl} = M_1 \delta_{ki}\delta_{lj} + M_8 t^a_{ki}t^a_{lj}
=\left(M_1-\frac{M_8}{2N_c}\right) \delta_{ki}\delta_{lj} +\frac{M_8}{2}\delta_{kj}\delta_{li} .\label{ampli}
\eeq
The singlet $M_1$ and octet $M_8$ contributions are from  the $Z$-boson fusion and  the  gluon-gluon fusion processes, respectively.  Their explicit forms are not important for this work. They can be found in the literature  \cite{Forshaw:2007vb}.  The $W$-boson fusion amplitude does not interfere with the above amplitude because $W$'s have an electric charge. As far as the color structure is concerned, the $W$-fusion process is identical to the $Z$-boson case, and does not require a separate consideration. 

\begin{figure}
  \includegraphics[width=1\linewidth]{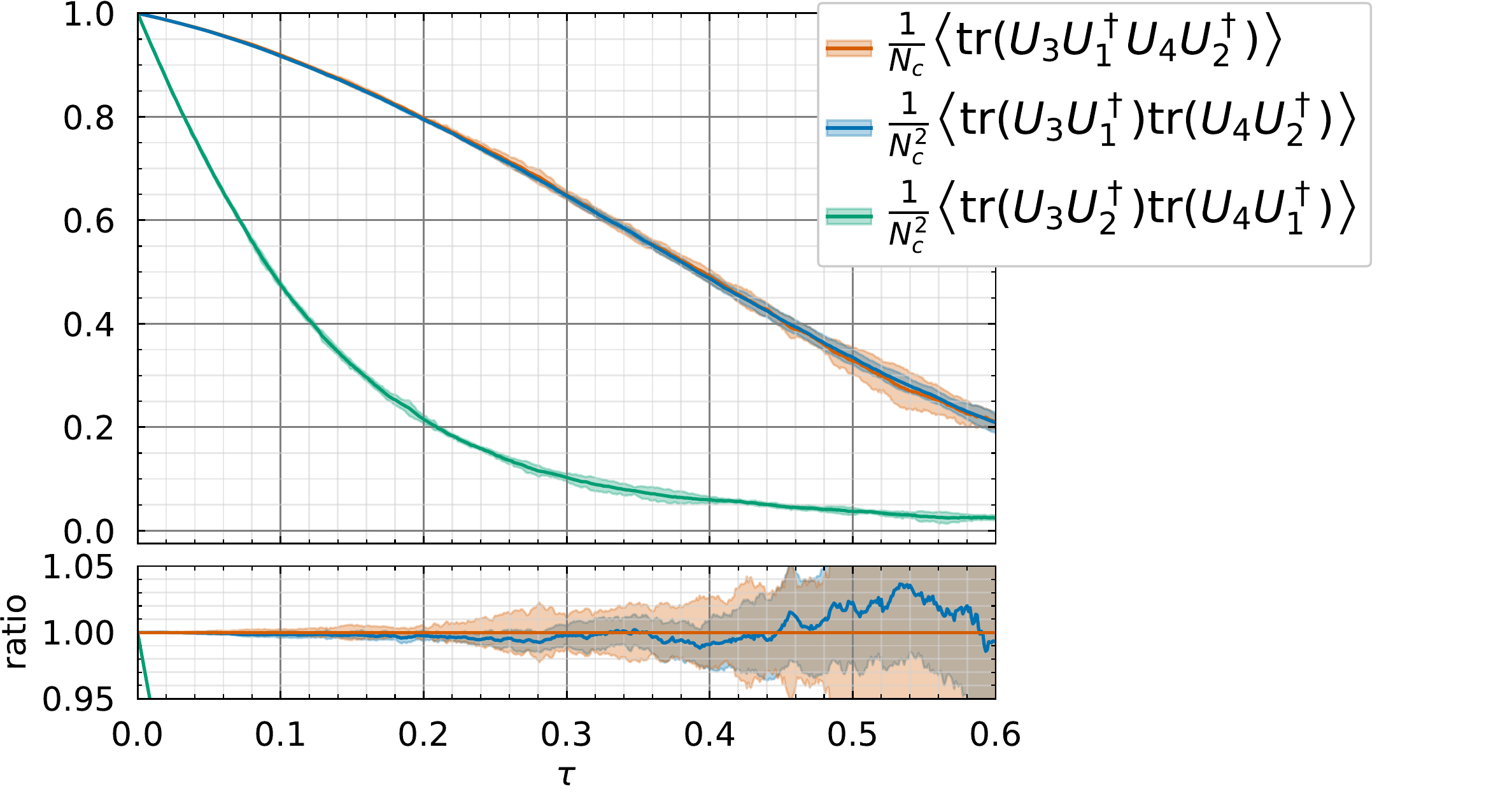}
\caption{Color multipoles relevant to the gap survival probability in $qq\to qqH$, see (\ref{in}). Only the real parts are plotted. }
\label{qq}
\end{figure}

We now dress up (\ref{ampli})
 by attaching soft gluons to external legs in the eikonal approximation. This converts (\ref{ampli}) into   
\beq
{\mathcal M}'_{ijkl} =\left(M_1-\frac{M_8}{2N_c}\right) (U_3U^\dagger_1)_{ki}(U_4U^\dagger_2)_{lj} +\frac{M_8}{2}(U_3U_2^\dagger)_{kj}(U_4U_1^\dagger)_{li}
\eeq
We then square it and average over $ij$, and sum over $kl$
\beq
\frac{1}{N_c^2} \sum_{ijkl}|{\mathcal M}'|^2 &=& \frac{1}{N_c^2}\left| M_1-\frac{M_8}{2N_c}\right|^2 \tr (U_3U^\dagger_1 V_1V_3^\dagger)
\tr (U_4U^\dagger_2 V_2V_4^\dagger) + \frac{|M_8|^2}{4N_c^2} \tr (U_3U_2^\dagger V_2V_3^\dagger) \tr(U_4U_1^\dagger V_1V_4^\dagger)\nonumber \\
&&+ \left(M_1-\frac{M_8}{2N_c}\right) \frac{M_8^*}{2N_c^2}\tr(U_3U_1^\dagger V_1V_4^\dagger U_4 U_2^\dagger V_2 V_3^\dagger) + (c.c.)
\eeq
Using the fact that $M_1$ and $M_8$ are real and relabeling $V^\dagger U \to U$ (see (\ref{pro})), we can write
\beq
\frac{1}{N_c^2} \sum_{ijkl}|{\mathcal M}'|^2 &=& \frac{1}{N_c^2}\left( M_1-\frac{M_8}{2N_c}\right)^2 \tr (U_3U^\dagger_1 )
\tr (U_4U^\dagger_2 ) + \frac{M_8^2}{4N_c^2} \tr (U_3U_2^\dagger) \tr(U_4U_1^\dagger )\nonumber \\
&&+  \left(M_1-\frac{M_8}{2N_c}\right)\frac{M_8}{N_c^2} \mathfrak{Re} \Bigl\{ \tr(U_3U_1^\dagger  U_4 U_2^\dagger ) \Bigr\}. \label{in}
\eeq
We see that the cross section involves  products of color dipoles ${\rm tr}(UU^\dagger){\rm tr}(UU^\dagger)$ and also a color quadrupole ${\rm tr}(UU^\dagger UU^\dagger)$. We evaluated these multipoles for a back-to-back configuration  $(\theta_3,\phi_3)= (\frac{\pi}{6},0)$ and $(\theta_4,\phi_4)=(\frac{5\pi}{6},\pi)$ with $\theta_{in}=\pi/3$.  As before, we average over 3000 random walks on $80\times 60$ and $60\times 40$  lattices.\footnote{Actually, the points $\theta_3=\pi/6=30^{{\rm o}}$ and $\theta_4=5\pi/6=150^{{\rm o}}$ are not exactly on a grid point of our lattices. To cope with this, we perform a linear interpolation (in $\cos\theta)$ of the results obtained for nearby grid points. On the $80\times 60$ lattice, we interpolate between  $\theta_3=29.14^{\rm o}$ and $31.99^{\rm o}$ (and similarly for $\theta_4$), and on the $60\times 40$ lattice,  $\theta_3=26.06^{\rm o}$ and $\theta_3=30.18^{\rm o}$. We do the same for all the plots below.} The results are plotted in Fig.~\ref{qq} with now \beq
\tau=\frac{\alpha_s}{\pi}\ln \frac{P_T}{E_{out}},
\eeq
where $P_T$ is the jet transverse momentum. 
Only the real parts  are plotted. The imaginary parts are consistent with zero within errors.

Surprisingly, we find, to a very good approximation,
\beq
\left\langle\frac{1}{N_c} {\rm tr}(U_3U_1^\dagger U_4 U_2^\dagger)\right\rangle \approx \left\langle \frac{1}{N_c^2}{\rm tr} (U_3U_1^\dagger) {\rm tr}(U_4U_2^\dagger) \right\rangle.
\label{fact}
\eeq
Namely, the color quadrupole factorizes into the product of color dipoles. We have checked that this property does not hold in each configuration, but  emerges only after averaging over many events. [Note  that (\ref{fact}) is trivially satisfied by the initial condition since $U=1$ everywhere.] It is surprising because such factorization has not been seen in the  previous studies of color multipoles  in the context of small-$x$ QCD, see e.g.,  \cite{Blaizot:2004wv,Dominguez:2011wm,Dumitru:2011vk,Shi:2017gcq}. The lesson learned in these studies is that a quadrupole does not factorize into dipoles in general. And when it factorizes in some limit and  in some sense, due to the cyclic property of trace, the two possible color singlet combinations ${\rm tr}(U_3U_1^\dagger) {\rm tr}(U_4U_2^\dagger)$ and ${\rm tr}(U_2^\dagger U_3) {\rm tr}(U_1^\dagger U_4)$  have to  appear symmetrically. However, in (\ref{fact}) only the former appears.  As shown in Fig.~\ref{qq}, the latter (green curve) is indeed numerically smaller since the dipoles have wider opening angles, but not negligibly smaller. While we do not understand the reason of this puzzling behavior, presumably it has to do with color coherence and angular ordering: Parton 3 prefers to pair up with parton 1 because then they can form a color-singlet dipole with a small opening angle ($\pi/6$ in this case) which is `protected' from the 2-4 dipole in the backward direction. What is striking about (\ref{fact}) is  that this tendency is pushed to the extreme. This point certainly deserves further studies. It is also interesting to see whether a relation analogous to (\ref{fact}) holds for any configuration of dipoles in the small-$x$ problem.  

\begin{figure}
  \centering
  \begin{subfigure}{0.49\linewidth}
    \centering
    \includegraphics[width=\textwidth]{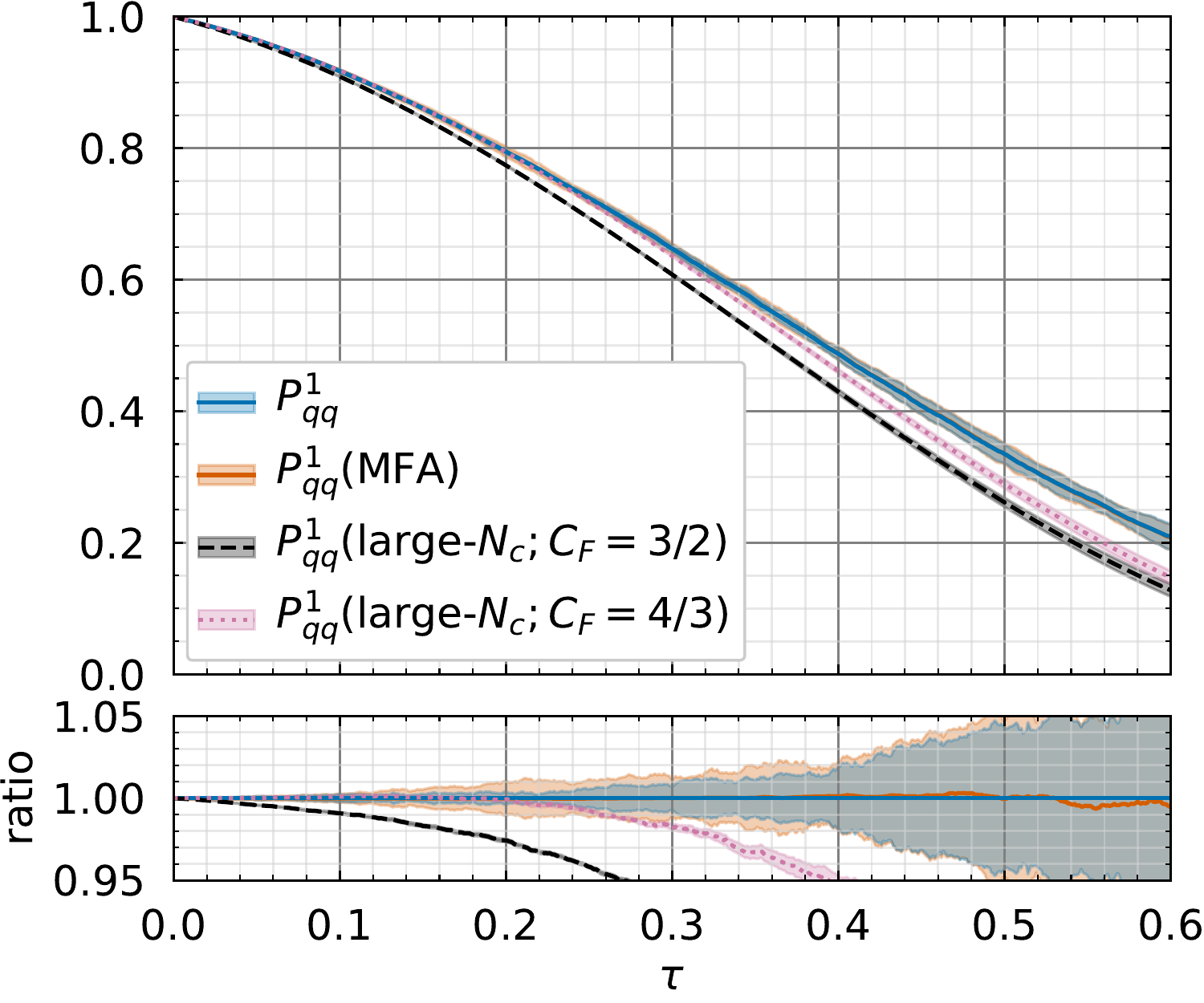}
    \caption{}
    \label{fig:Pqq1}
  \end{subfigure}
  \hfill
  \begin{subfigure}{0.49\linewidth}
    \centering
    \includegraphics[width=\linewidth]{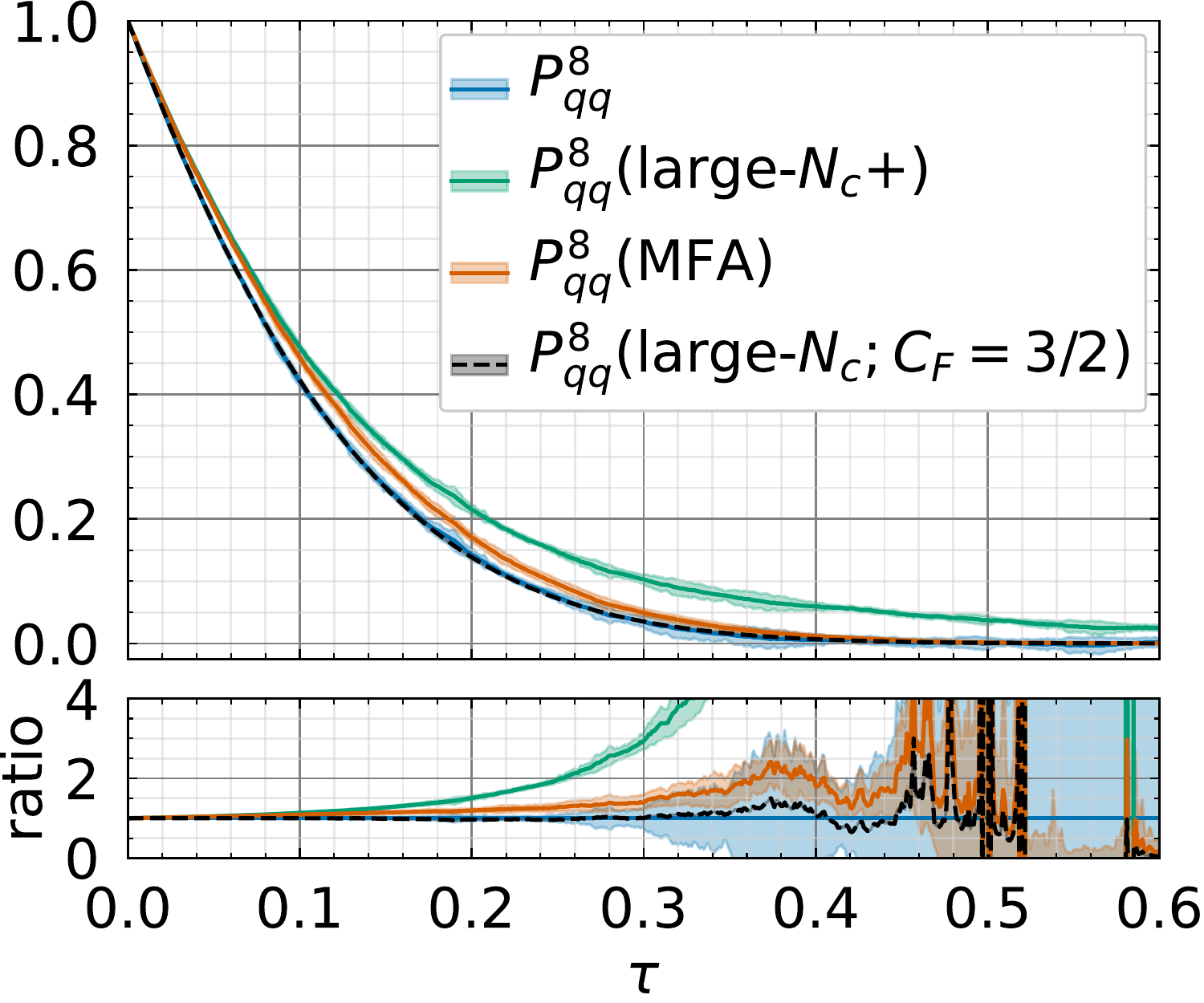}
    \caption{}
    \label{fig:Pqq8}
  \end{subfigure}
  \caption{Gap survival probabilities in $qq\to qqH$, color-singlet (a) and color-octet (b) channels.}
\end{figure}

Let us now consider the implications of (\ref{fact}). We immediately notice that if we use (\ref{fact}) in (\ref{in}), the interference term $\propto M_1M_8$ between the $Z$-boson and gluon fusion amplitudes vanishes. Actually, that the interference effect is numerically very small was already observed in \cite{Forshaw:2007vb}. Even without soft gluon emissions,  it is already suppressed at the tree level due to a cancellation between contributions from different flavors and helicities. [$M_1$ depends on these quantum numbers.]  Interestingly, in addition to this `accidental' suppression, here we find another  dynamical source of suppression which makes the interference term really small.  
After using the relation (\ref{fact}) in (\ref{in}), we get 
\beq
 M_1^2P_{qq}^1 + \frac{N_c^2-1}{4N_c^2}M_8^2P_{qq}^8, \eeq
where the probabilities 
\beq
P_{qq}^1&\equiv& \frac{1}{N_c^2} \left\langle {\rm tr}(U_3U_1^\dagger){\rm tr}(U_4U_2^\dagger)\right\rangle , \\
P^8_{qq}&\equiv& \frac{1}{N_c^2-1} \left\langle {\rm tr}(U_3 U_2^\dagger) {\rm tr} (U_4 U_1^\dagger) - \frac{1}{N_c^2}  {\rm tr}(U_3U_1^\dagger){\rm tr}(U_4U_2^\dagger) %
\right\rangle, \label{plot}
\eeq
are normalized to unity at $\tau=0$. In Fig.~\ref{fig:Pqq1}, we plot $P_{qq}^1$ together with its mean-field approximated version 
\beq
P^1_{qq}({\rm MFA}) \equiv \left\langle \frac{1}{N_c}{\rm tr}(U_3U_1^\dagger)\right\rangle \left\langle \frac{1}{N_c}{\rm tr}(U_4U_2^\dagger)\right\rangle.
\eeq
as well as the large-$N_c$ version
\beq
P^1_{qq}(\text{large-}N_c)\equiv P_{13}P_{24}
\eeq
where $P_{\alpha\beta}$ is the solution of the BMS equation (\ref{bms}) with $C_F=\frac{N_c^2-1}{2N_c}=\frac{4}{3}$ (pink curve) and $C_F\approx \frac{N_c}{2}=\frac{3}{2}$ (black curve). 

In Fig.~\ref{fig:Pqq8}, we plot $P_{qq}^8$ and its variants 
\beq
P_{qq}^{8}(\text{large-}N_c+)&\equiv& \frac{1}{N_c^2} \left\langle {\rm tr}(U_3 U_2^\dagger) {\rm tr} (U_4 U_1^\dagger)\right\rangle,  \label{one}\\
P_{qq}^8({\rm MFA})&\equiv& \left\langle \frac{1}{N_c}{\rm tr}(U_3U_2^\dagger)\right\rangle \left\langle \frac{1}{N_c}{\rm tr}(U_4U_1^\dagger)\right\rangle \label{two} \\
P_{qq}^8(\text{large-}N_c)&\equiv& P_{23}P_{14}\label{bo}
\eeq
(\ref{one}) is obtained from $P_{qq}^8$ by keeping terms with the largest power of $N_c$ under the counting rules ${\rm tr}(UU^\dagger)\sim {\rm tr}(UU^\dagger UU^\dagger)\sim N_c$, $N_c^2-1 \approx N_c^2$. For the lack of a better name, we refer to it as the `large-$N_c+$' approximation, although it is a bit misleading since  we evaluate the resulting expression fully at $N_c=3$. The `genuine' large-$N_c$ approximation is given by (\ref{bo}) where $P_{\alpha\beta}$ is calculated from the BMS equation with $C_F=N_c/2$. 

We immediately notice that the MFA $\langle AB\rangle \approx \langle A\rangle\langle B\rangle$ holds almost perfectly in Fig.~\ref{fig:Pqq1} but fails completely in Fig.~\ref{fig:Pqq8} for $\tau \gtrsim 0.2$ (compare the  green  and orange  curves). Combining with the previous example  Fig.~\ref{fig:2b}, we can infer that the MFA is good when the two dipoles are far apart  in solid angles, but violated when they are close to each other, see  Fig.~\ref{2to2}(right). This supports our previous claim that  the breakdown of the MFA is due to the spatial correlation among soft gluons which gets stronger when they are close to each other \cite{Hatta:2007fg,Avsar:2008ph}.

We next observe that, surprisingly, the full result (\ref{plot}) (blue curve in Fig.~\ref{fig:Pqq8}) agrees almost perfectly with the large-$N_c$ result (\ref{bo}).   This is similar to the relation $P_H\approx P(\text{large-}N_c)$ found in the previous section, but unlike there, this time we do not have a simple explanation. [Note, however, that (\ref{plot}) reduces to (\ref{higg}) in the limits $\theta_3\to 0$, $\theta_4\to \pi$.] Our result indicates that the large suppression factor when going from (\ref{one}) to (\ref{two}) perfectly mimics the second term of (\ref{plot}) discarded in  the large-$N_c$ approximation.
Despite the explicit factor of $1/N_c^2$, the second term is not at all negligible compared to the first term when $\tau\gtrsim 0.2$. In fact,  $P_{qq}^8$ vanishes around $\tau \sim 0.4$ due to an almost  exact cancellation between the two terms. How can $P_{qq}^8(\text{large-}N_c)$ know about this delicate cancellation when it totally ignores the 1-3 and 2-4 dipoles? An easy explanation is that the agreement is just an accident, but there may be a deep reason. We shall return to this issue later.  

Finally, the large-$N_c$ approximation is  violated in the singlet sector $P_{qq}^1$ (compare the blue and black curves in Fig.~\ref{fig:Pqq1}). Actually, from our experience in \cite{Hatta:2013iba}, we  expected the factorized product  $P_{qq}^1(\text{MFA})$ to be very close to the solution of the BMS equation with $C_F=4/3$ (pink curve), but we see a clear deviation for $\tau>0.3$. While this  may be physical, one has to be very careful about lattice artifacts. The 1-3 and 2-4 dipoles involved in the singlet channel have a small opening angle, and hence they may be more susceptible to lattice discretization  errors.\footnote{We thank Gavin Salam for pointing this out.} Besides, such errors are doubled when computing the square $P_{qq}^1\sim (P_{13})^2$.  To settle this issue, we need simulations on much finer lattices, which is however computationally challenging in the present approach.

All these results are  in stark  contrast to the case of $e^+e^-$ annihilation where one does not see any unusual behavior at such early `times' $\tau\sim 0.2$ \cite{Hatta:2013iba,Hagiwara:2015bia}. 
In particular, the finite-$N_c$ corrections to the color dipole  is quite small in this regime. In the corresponding small-$x$ problem, it has even been argued  that the $1/N_c^2$ corrections  are smaller by orders of magnitude than the naive expectation $1/N_c^2\sim 10$\% \cite{Kovchegov:2008mk}. However, in hadron-hadron collisions,  the gap survival probability $P_{hh}$ consists of higher multipoles and becomes small, say $P_{hh}< 0.2$, already in the phenomenologically relevant region of $\tau$. In this region, naively subdominant effects (spatial correlations, finite-$N_c$) can give  corrections of order unity. Barring further ``accidents" to happen, it is simply best to avoid any  approximations  under such circumstances.

\begin{figure}
  \includegraphics[width=0.7\linewidth]{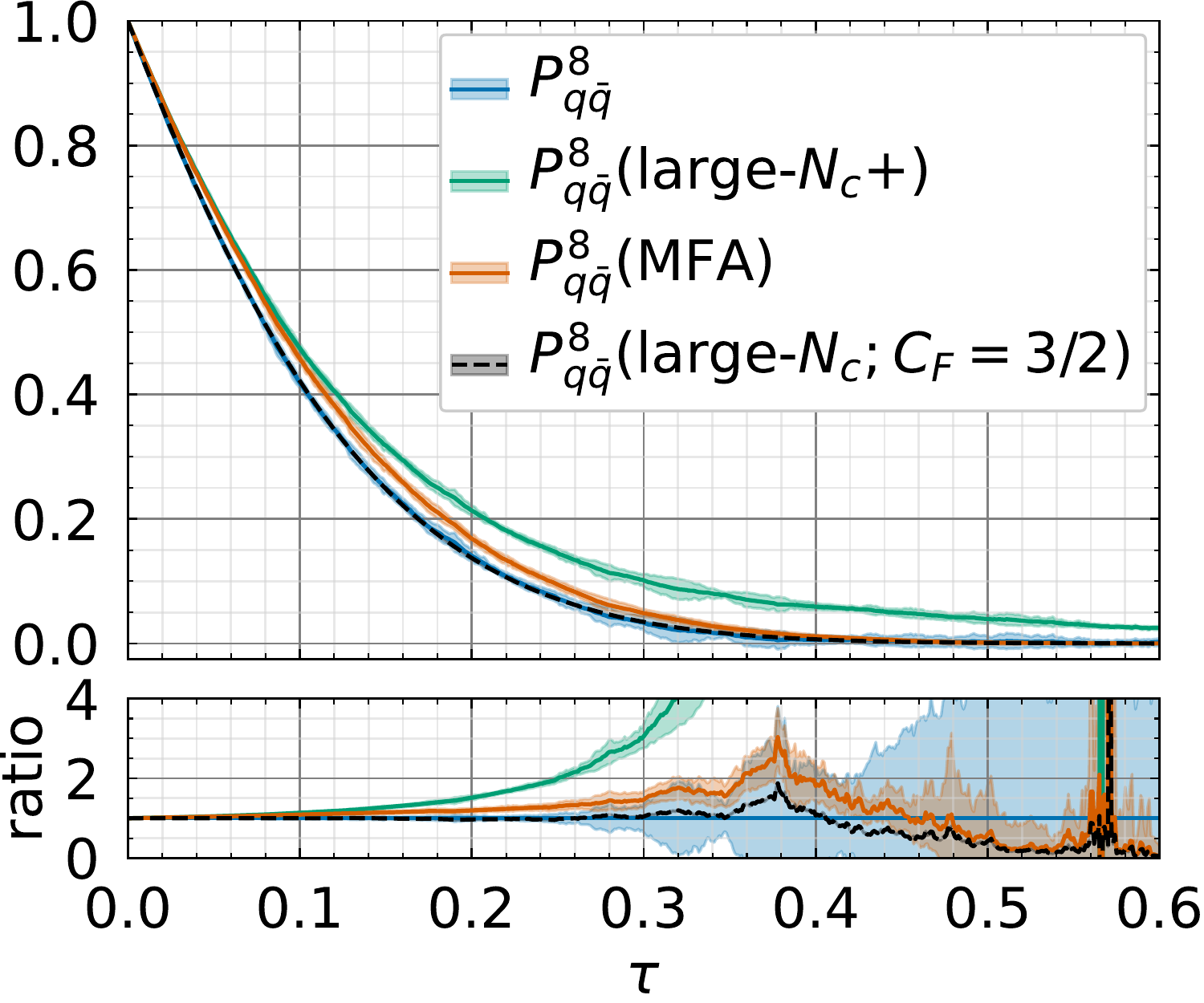}
  \caption{Gap survival probability in  $q\bar{q}\to q\bar{q}$, octet channel (\ref{qqbar}). }
  \label{fig:mean3421}
\end{figure}

\subsection{\texorpdfstring{$q\bar{q}\to q\bar{q}H$}{q qbar -> q qbar H}}

Next consider the channel $q\bar{q}\to q\bar{q}H$.  
In the eikonal approximation $\bar{v}(p_2)\gamma^\mu v(p_4)\approx 2p^\mu_2$, we write
\beq
{\mathcal M}_{ijkl} = -M_1 \delta_{ki}\delta_{lj} -M_8  t^a_{ki}t^a_{jl} = -\left( M_1 -\frac{M_8}{2N_c} \right) \delta_{ki}\delta_{lj} -\frac{M_8}{2}\delta_{kl}\delta_{ij},
\nonumber \\
\to {\mathcal M}_{ijkl}'= -\left( M_1 -\frac{M_8}{2N_c} \right) (U_3U_1^\dagger)_{ki}(U_2U_4^\dagger)_{jl} -\frac{M_8}{2}(U_3U_4^\dagger)_{kl}(U_2U_1^\dagger)_{ji}.
\eeq
A minus sign is needed when replacing a quark with an antiquark. 
Squaring and averaging over color indices, we get
\beq
\frac{1}{N_c^2}\sum_{ijkl}|{\mathcal M}'|^2 = \frac{1}{N_c^2}\left(M_1 -\frac{M_8}{2N_c} \right)^2 \tr (U_3U_1^\dagger)\tr(U_2U_4^\dagger)
+\frac{M_8^2}{4N_c^2} \tr(U_3U_4^\dagger) \tr(U_2U_1^\dagger)
\nonumber \\
+  \left( M_1 -\frac{M_8}{2N_c} \right)  \frac{M_8}{N_c^2} \mathfrak{Re} \Bigl\{ \tr(U_3U_1^\dagger U_2U_4^\dagger) \Bigr\}. \label{check}
\eeq
We have checked that, similarly to (\ref{fact}), 
\beq
\left\langle\frac{1}{N_c} {\rm tr}(U_3U_1^\dagger U_2 U_4^\dagger)\right\rangle \approx \left\langle \frac{1}{N_c^2}{\rm tr} (U_3U_1^\dagger) {\rm tr}(U_2U_4^\dagger) \right\rangle \approx \left\langle\frac{1}{N_c} {\rm tr}(U_3U_1^\dagger U_4 U_2^\dagger)\right\rangle,
\eeq
so that again the interference term $\sim M_1M_8$ in (\ref{check}) is negligibly small. %
 Eq.~(\ref{check}) then reduces to  
\beq
M_1^2P_{q\bar{q}}^1 + \frac{N_c^2-1}{4N_c^2}M_8^2 P^8_{q\bar{q}}
\eeq
where  $P_{q\bar{q}}^1 \approx P_{qq}^1$ and 
\beq
P^8_{q\bar{q}} \equiv \frac{1}{N_c^2-1}\left\langle {\rm tr}(U_3U_4^\dagger){\rm tr}(U_2U_1^\dagger) -\frac{1}{N_c^2}{\rm tr}(U_3U_1^\dagger){\rm tr}(U_2U_4^\dagger)\right\rangle. \label{qqbar}
\eeq 
This is plotted in Fig.~\ref{fig:mean3421} together with its three variants
\beq
P^8_{q\bar{q}}(\text{large-}N_c+) &\equiv& \frac{1}{N_c^2}\left\langle {\rm tr}(U_3U_4^\dagger){\rm tr}(U_2U_1^\dagger)\right\rangle, \label{qqbar1} \\
P^8_{q\bar{q}}({\rm MFA}) &\equiv& \left\langle \frac{1}{N_c}{\rm tr}(U_3U_4^\dagger)\right\rangle \left\langle \frac{1}{N_c}{\rm tr}(U_2U_1^\dagger)\right\rangle \label{qqbar2}\\
P_{q\bar{q}}^8(\text{large-}N_c;C_F=3/2)&\equiv & P_{34}P_{12} \label{largeqqbar}
\eeq
As expected, the MFA does not hold for  $\tau\gtrsim 0.2$ because the 3-4 and 1-2 dipoles are close in angles, see Fig.~\ref{2to2}(right). Again  the full result (\ref{qqbar}) agrees almost perfectly with the large-$N_c$ result (\ref{largeqqbar}) despite a series of approximations  (\ref{qqbar})$\to$(\ref{qqbar1})$\to$(\ref{qqbar2})$\to$(\ref{largeqqbar}) involved. 
Although nontrivial, this  may not come as an additional surprise  since the two processes $qq\to qq$ and $q\bar{q}\to q\bar{q}$ are rather similar in the present setup.

\subsection{\texorpdfstring{$qg\to qgH$, $\bar{q}g \to \bar{q}gH$}{q g -> q g H, qbar g -> qbar g H}}

\begin{figure}
  \includegraphics[width=0.7\linewidth]{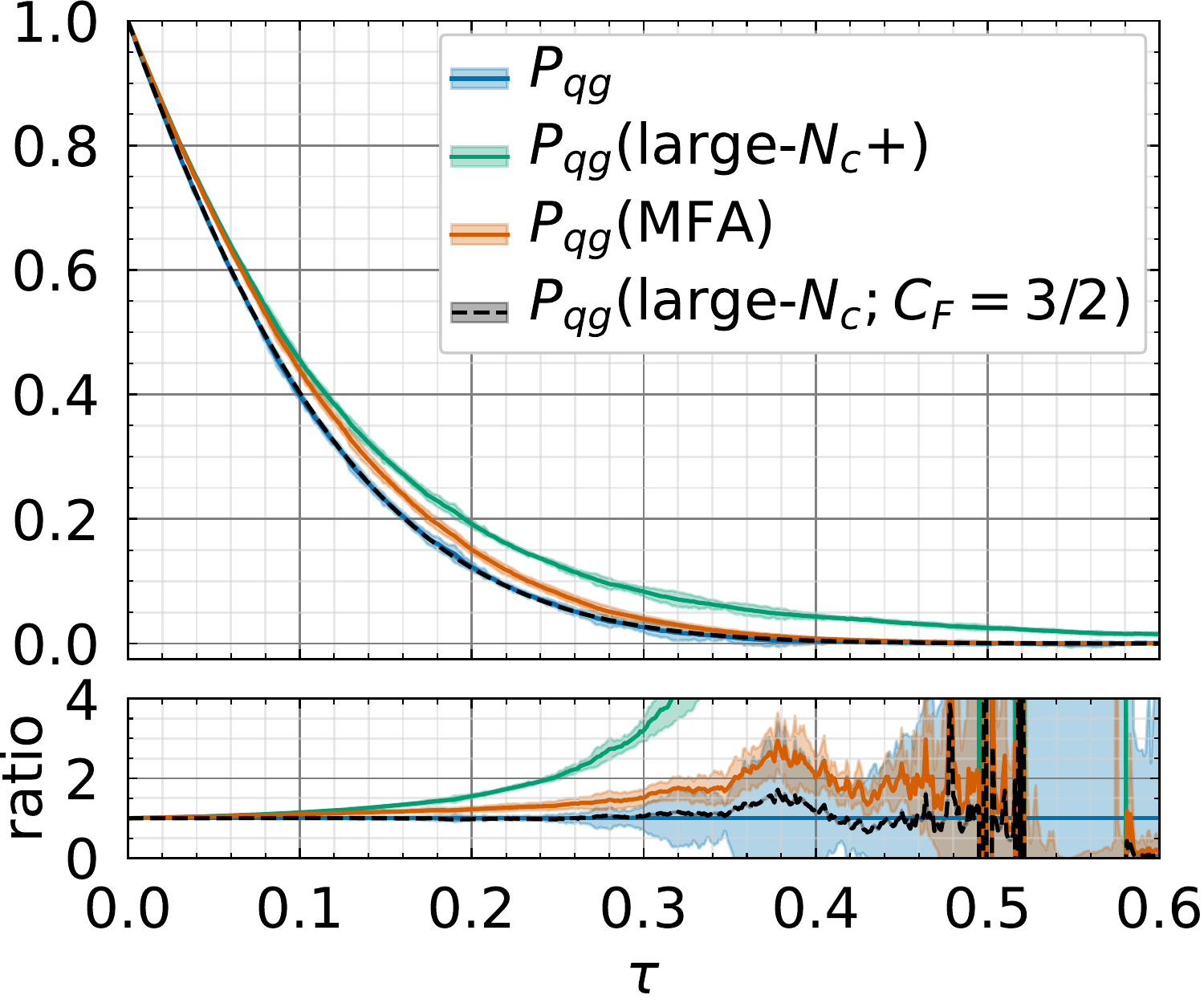}
  \caption{Gap survival probability in the $qg\to qg$ channel.}
  \label{fig:qgh}
\end{figure}

Next we turn to the case which involves a gluon in the initial state $qg\to qgH$. ($\bar{q}g \to \bar{q}gH$ is entirely analogous and will be omitted.)  
There is no vector boson fusion contribution in this channel. The amplitude is given by
\beq
{\mathcal M}^{ab}_{ik}= t^c_{ki} T^c_{ba} M_8 %
\quad \to \quad  
{\mathcal M}^{\prime ab}_{ik} = (U_3 t^c U^\dagger_1)_{ki} (\tilde{U}_4 T^c \tilde{U}_2^\dagger)_{ba}M_8, %
\eeq
where $a,b$ are the color indices of the initial and final gluons with $T^c_{ba}=-if_{cba}$. 
Squaring and averaging over color indices, we get 
\beq
\frac{1}{N_c(N_c^2-1)}\sum_{ikab} |{\mathcal M}^\prime|^2 &=&\frac{M_8^2}{N_c(N_c^2-1)} \tr(U_3 t^c U_1^\dagger V_1 t^d V_3^\dagger) \tr(\tilde{U_4}T^c \tilde{U}_2^\dagger \tilde{V}_2 T^d \tilde{V}^\dagger_4) \nonumber \\
&\to& \frac{M_8^2}{N_c(N_c^2-1)} \tr (t^c U^\dagger_1 t^d U_3) \tr ( T^c \tilde{U}^\dagger_2 T^d \tilde{U}_4) \nonumber \\
&=& \frac{M_8^2}{4N_c(N_c^2-1)} \biggl( \tr (U_2 U^\dagger_1)  \tr (U_3 U_4^\dagger) \tr (U_4 U_2^\dagger) + \tr (U_4 U_1^\dagger) \tr (U_2 U_4^\dagger) \tr (U_3 U_2^\dagger) \nonumber \\
&& \qquad \qquad -\tr (U_4 U_2^\dagger U_3 U_4^\dagger U_2 U_1^\dagger) - \tr (U_4 U_1^\dagger U_2 U_4^\dagger U_3 U_2^\dagger) \biggr) \nn
&& \equiv \frac{M_8^2}{2} P_{qg} \label{gg}
\eeq
We now have  color sextupoles ${\rm tr}(UU^\dagger UU^\dagger UU^\dagger)$ as well as products of three dipoles ${\rm tr}(UU^\dagger){\rm tr}(UU^\dagger) {\rm tr}(UU^\dagger)$.  
 $P_{qg}$ is plotted in Fig.~\ref{fig:qgh} together with its  approximated versions
 \beq
&& P_{qg}(\text{large-}N_c+) \equiv  \frac{1}{2N_c^3} \left\langle \tr (U_2 U^\dagger_1)  \tr (U_3 U_4^\dagger) \tr (U_4 U_2^\dagger) + \tr (U_4 U_1^\dagger) \tr (U_2 U_4^\dagger) \tr (U_3 U_2^\dagger)  \right\rangle \\
 && P_{qg}({\rm MFA}) \equiv  \frac{1}{2}\left\langle \frac{{\rm tr}(U_2U_1^\dagger)}{N_c}\right\rangle \left\langle \frac{{\rm tr}(U_3U_4^\dagger)}{N_c}\right\rangle \left\langle \frac{{\rm tr}(U_4U_2^\dagger)}{N_c}\right\rangle  \nn && \qquad \qquad \qquad \qquad 
 +\frac{1}{2}\left\langle \frac{{\rm tr}(U_4U_1^\dagger)}{N_c}\right\rangle \left\langle\frac{{\rm tr}(U_2U_4^\dagger)}{N_c}\right\rangle \left\langle \frac{{\rm tr}(U_3U_2^\dagger)}{N_c}\right\rangle \\
 && P_{qg}(\text{large-}N_c;C_F=3/2)\equiv  \frac{1}{2}(P_{12}P_{34}+P_{14}P_{23})P_{24}. \label{largegq}
 \eeq
 {\it A priori}, one would expect $P^8_{qq}\approx  P^8_{q\bar{q}}>P_{qg}$, but their difference turns out to be numerically rather small. Fig.~\ref{fig:qgh} looks almost the same as the previous plots. 
 This is because, roughly,
 \beq
P_{qg} \sim P_{41}P_{24}P_{32}\sim P_{24}P^8_{qq} \label{13}
\eeq
 and $P_{24}$ is of order unity  (for example, $P_{24}\approx 0.8$ when $\tau=0.3$). For the first time, we observe the large violation of the MFA in the three-dipole sector (green versus orange curves).  The color sextupoles  in (\ref{gg}) are nominally subleading $\sim{\cal O}(1/N_c^2)$ in the $N_c$ counting, but they  
completely cancel the leading-$N_c$ terms for $\tau\gtrsim 0.3$. After this cancellation,  once again, the exact result is very close to the large-$N_c$ result (\ref{largegq})!

\subsection{\texorpdfstring{$gg\to ggH$}{g g -> g g H}}
\begin{figure}
  \includegraphics[width=0.7\linewidth]{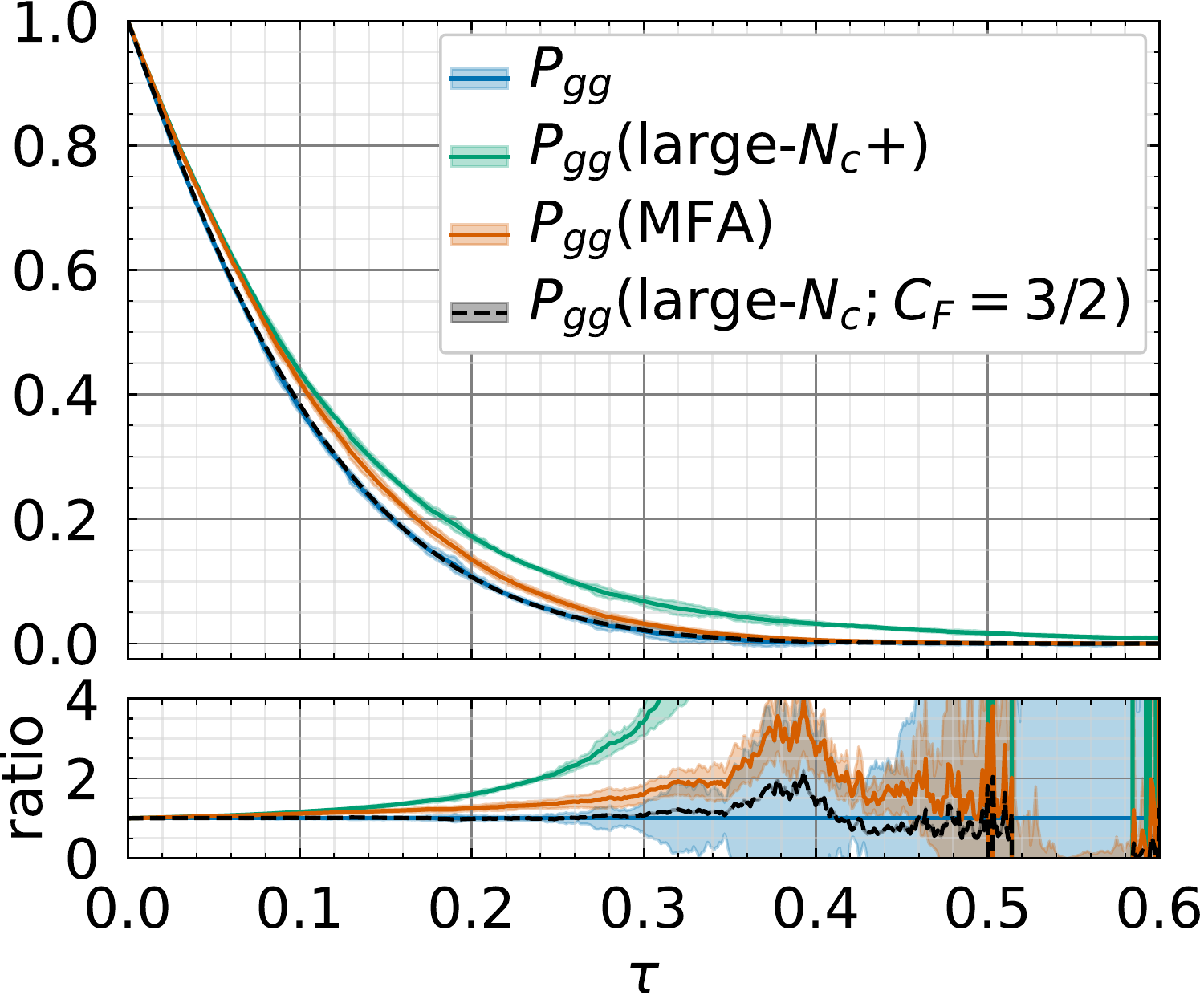}
  \caption{Gap survival probability in the $gg\to gg$ channel. }
  \label{fig:ggh}
\end{figure}

Finally, the amplitude in the $gg\to gg$ channel is
\beq
{\mathcal M}_{aa'bb'} =M_8 T^c_{ba}T^c_{b'a'}  \quad \to \quad {\mathcal M}'_{aa'bb'}= M_8 (\tilde{U}_3T^c\tilde{U}_1^\dagger)_{ba}
(\tilde{U}_4T^c\tilde{U}_2)_{b'a'}
\eeq
Proceeding as before, we find 
\beq
&&  \frac{1}{(N_c^2-1)^2}\sum_{aa'bb'}|{\mathcal M}^\prime|^2 \nonumber \\
&&\to \frac{M_8^2}{(N_c^2-1)^2} \tr (T^c \tilde{U}^\dagger_1 T^d \tilde{U}_3) \tr ( T^c \tilde{U}^\dagger_2 T^d \tilde{U}_4) \nonumber \\
&&= \frac{M_8^2}{2(N_c^2-1)^2}\mathfrak{Re} \Biggl\{\tr (U_1 U_3^\dagger)\left( \tr (U_2U_4^\dagger) \tr (U_4 U_1^\dagger) \tr (U_3 U_2^\dagger) +  \tr (U_4 U_2^\dagger) \tr (U_2 U_1^\dagger) \tr (U_3 U_4^\dagger) \right) \nonumber \\
&& \qquad \qquad \qquad  \qquad -\tr (U_1U_3^\dagger) \left( \tr (U_4U_2^\dagger U_3 U_4^\dagger U_2 U_1^\dagger)
+\tr (U_2 U_4^\dagger U_3 U_2^\dagger U_4 U_1^\dagger) \right) \nonumber \\
&& \qquad \qquad \qquad \qquad -\tr (U_2U_4^\dagger) \left( \tr (U_1U_3^\dagger U_4 U_1^\dagger U_3 U_2^\dagger)
+\tr (U_3 U_1^\dagger U_4 U_3^\dagger U_1 U_2^\dagger) \right) \nonumber \\
&& \qquad \qquad \qquad \qquad +  \tr (U_3U_2^\dagger U_4 U_1^\dagger ) \tr (U_2 U_4^\dagger U_1 U_3^\dagger) + \tr (U_1U_2^\dagger U_4 U_3^\dagger) \tr (U_2 U_4^\dagger U_3U_1^\dagger)  \Biggr\} \nn 
&& \equiv \frac{N_c^2}{N_c^2-1} M_8^2 P_{gg}.
\eeq
This features various color multipoles consisting of eight Wilson lines. [Note that it does not contain `color octupoles' ${\rm tr}(UU^\dagger UU^\dagger UU^\dagger UU^\dagger)$.] $P_{gg}$ is plotted in Fig.~\ref{fig:ggh} together with its three approximations
\beq
&&P_{gg}(\text{large-}N_c+) \equiv  
\frac{1}{2N_c^4} \left\langle \tr (U_1 U_3^\dagger)\left( \tr (U_2U_4^\dagger) \tr (U_4 U_1^\dagger) \tr (U_3 U_2^\dagger) +  \tr (U_4 U_2^\dagger) \tr (U_2 U_1^\dagger) \tr (U_3 U_4^\dagger)\right) \right\rangle \\ 
&&P_{gg}({\rm MFA}) \equiv  \frac{1}{2}\left\langle \frac{{\rm tr}(U_1U^\dagger_3)}{N_c}\right\rangle  \Biggl\{ \left\langle \frac{{\rm tr}(U_2U^\dagger_4)}{N_c}\right\rangle   \left\langle \frac{{\rm tr}(U_4U^\dagger_1)}{N_c}\right\rangle  \left\langle \frac{{\rm tr}(U_3U^\dagger_2)}{N_c}\right\rangle \nn 
&& \qquad \qquad \qquad \qquad + \left\langle \frac{{\rm tr}(U_4U^\dagger_2)}{N_c}\right\rangle  \left\langle \frac{{\rm tr}(U_2U^\dagger_1)}{N_c}\right\rangle  \left\langle \frac{{\rm tr}(U_3U^\dagger_4)}{N_c}\right\rangle  \Biggr\} \\
&&P_{gg}(\text{large-}N_c;C_F=3/2)\equiv  \frac{1}{2}P_{13}P_{24}(P_{14}P_{23}+P_{12}P_{34}). \label{largegg}
\eeq 
By now the pattern is routine.  The MFA is violated also in the four-dipole sector, and the terms that consist of higher multipoles  are not negligible compared to the leading-$N_c$, dipole terms. Nevertheless, the final result is very well approximated by the large-$N_c$ result (\ref{largegg}) with $C_F=3/2$. \\

At this point, we must abandon the idea  that the agreement between the full-$N_c$ and large-$N_c$  results  is  accidental. Actually,  we have also tried asymmetric jets configurations such as $(\theta_{3},\theta_4)=(15^{\rm o}, 150^{\rm o})$ and $(45^{\rm o},150^{\rm o}$) and arrived at the same conclusion. This is a striking observation.  We would have expected that  $P(\text{large-}N_c)$ would be the worst approximation of all. Indeed, the naively subleading terms in $N_c$ are numerically significant and push the green curve down to the blue curve.  Yet, the large-$N_c$ approximation somehow `knows' this cancellation in advance, and gives  almost correct results in terms of most simplistic formulas. How this is possible is unclear to us. 

\section{Conclusions}

In this paper, we have performed the  resummation of leading non-global logarithms   for two specific observables in proton-proton collisions at the LHC\@. No approximation is used for the number of colors $N_c=3$. In contrast to $e^+e^-$ annihilation studied previously, higher order color multipoles  come into play. They can be straightforwardly evaluated in our formalism developed in \cite{Hatta:2013iba}. Our  simulations have revealed several surprising features, such as the reduction of a  quadrupole into the product of dipoles (\ref{fact}) and the failure of the `large-$N_c+$' and mean field approximations.  In particular, terms naively  subleading in $N_c$ can completely cancel the leading-$N_c$ terms when $\tau\gtrsim 0.3$. 

The final surprise is that, despite  these highly nontrivial finite-$N_c$ effects, the large-$N_c$ result $P(\text{large-}N_c)$ with $C_F=3/2$, which naively appears to be the least precise approximation, agrees perfectly with the  exact $N_c=3$ result at least up to $\tau\sim 0.3$. We have confirmed this in all the subprocesses studied in this paper except in  the singlet channel $P_{qq}^1$. In $H\to gg$, there is a semi-analytical  explanation of how this might occur (see the paragraph below (\ref{sud})), but for the dijet case, at the moment we do not understand why this should be the case.  A close inspection of the leading order Sudakov  ${\cal O}(\alpha_s)$  and non-global    ${\cal O}(\alpha_s^2)$ logarithms in $2\to 2$ processes may help resolve this issue. 

It remains to be seen whether a similar conclusion holds for other observables. Even for the dijet problem, a number of tests can be carried out. For example, one  can relax the eikonal approximation used to derive (\ref{ampli}), cf., \cite{Hatta:2013qj}, or one can use  different definitions of the `out' region.  We leave this to future work. 
If, after all these tests, the relation $P(N_c=3)\approx P(\text{large-}N_c)$ turns out to be robust, it is good news because  one can approximately get full-$N_c$ results in hadron collisions using the known large-$N_c$ frameworks \cite{Dasgupta:2001sh,Dasgupta:2002bw,Banfi:2002hw}.

\section*{Acknowledgments}
We are grateful to Gavin Salam and Gregory Soyez for encouragement and many useful discussions on various topics including their new results in \cite{1831912}. We also thank  Bowen Xiao for correspondence.  Y.~H. thanks the Yukawa Institute for Theoretical Physics, Kyoto University for hospitality. 
The work by T.~U. is in part supported by JSPS KAKENHI Grant Number 19K03831. The work by Y.~H. is supported by the U.S. Department of Energy, Office of Science, Office of Nuclear Physics, under contract number DE- SC0012704, and also by  Laboratory Directed Research and Development (LDRD) funds from Brookhaven Science Associates.

\bibliography{references}

\begin{thebibliography}{39}%
\makeatletter
\providecommand \@ifxundefined [1]{%
 \@ifx{#1\undefined}
}%
\providecommand \@ifnum [1]{%
 \ifnum #1\expandafter \@firstoftwo
 \else \expandafter \@secondoftwo
 \fi
}%
\providecommand \@ifx [1]{%
 \ifx #1\expandafter \@firstoftwo
 \else \expandafter \@secondoftwo
 \fi
}%
\providecommand \natexlab [1]{#1}%
\providecommand \enquote  [1]{``#1''}%
\providecommand \bibnamefont  [1]{#1}%
\providecommand \bibfnamefont [1]{#1}%
\providecommand \citenamefont [1]{#1}%
\providecommand \href@noop [0]{\@secondoftwo}%
\providecommand \href [0]{\begingroup \@sanitize@url \@href}%
\providecommand \@href[1]{\@@startlink{#1}\@@href}%
\providecommand \@@href[1]{\endgroup#1\@@endlink}%
\providecommand \@sanitize@url [0]{\catcode `\\12\catcode `\$12\catcode
  `\&12\catcode `\#12\catcode `\^12\catcode `\_12\catcode `\%12\relax}%
\providecommand \@@startlink[1]{}%
\providecommand \@@endlink[0]{}%
\providecommand \url  [0]{\begingroup\@sanitize@url \@url }%
\providecommand \@url [1]{\endgroup\@href {#1}{\urlprefix }}%
\providecommand \urlprefix  [0]{URL }%
\providecommand \Eprint [0]{\href }%
\providecommand \doibase [0]{http://dx.doi.org/}%
\providecommand \selectlanguage [0]{\@gobble}%
\providecommand \bibinfo  [0]{\@secondoftwo}%
\providecommand \bibfield  [0]{\@secondoftwo}%
\providecommand \translation [1]{[#1]}%
\providecommand \BibitemOpen [0]{}%
\providecommand \bibitemStop [0]{}%
\providecommand \bibitemNoStop [0]{.\EOS\space}%
\providecommand \EOS [0]{\spacefactor3000\relax}%
\providecommand \BibitemShut  [1]{\csname bibitem#1\endcsname}%
\let\auto@bib@innerbib\@empty
%</preamble>
\bibitem [{\citenamefont {Platzer}\ and\ \citenamefont
  {Sjodahl}(2012)}]{Platzer:2012np}%
  \BibitemOpen
  \bibfield  {author} {\bibinfo {author} {\bibfnamefont {S.}~\bibnamefont
  {Platzer}}\ and\ \bibinfo {author} {\bibfnamefont {M.}~\bibnamefont
  {Sjodahl}},\ }\href {\doibase 10.1007/JHEP07(2012)042} {\bibfield  {journal}
  {\bibinfo  {journal} {JHEP}\ }\textbf {\bibinfo {volume} {07}},\ \bibinfo
  {pages} {042} (\bibinfo {year} {2012})},\ \Eprint
  {http://arxiv.org/abs/1201.0260} {arXiv:1201.0260 [hep-ph]} \BibitemShut
  {NoStop}%
\bibitem [{\citenamefont {Nagy}\ and\ \citenamefont
  {Soper}(2015)}]{Nagy:2015hwa}%
  \BibitemOpen
  \bibfield  {author} {\bibinfo {author} {\bibfnamefont {Z.}~\bibnamefont
  {Nagy}}\ and\ \bibinfo {author} {\bibfnamefont {D.~E.}\ \bibnamefont
  {Soper}},\ }\href {\doibase 10.1007/JHEP07(2015)119} {\bibfield  {journal}
  {\bibinfo  {journal} {JHEP}\ }\textbf {\bibinfo {volume} {07}},\ \bibinfo
  {pages} {119} (\bibinfo {year} {2015})},\ \Eprint
  {http://arxiv.org/abs/1501.00778} {arXiv:1501.00778 [hep-ph]} \BibitemShut
  {NoStop}%
\bibitem [{\citenamefont {Dasgupta}\ \emph {et~al.}(2018)\citenamefont
  {Dasgupta}, \citenamefont {Dreyer}, \citenamefont {Hamilton}, \citenamefont
  {Monni},\ and\ \citenamefont {Salam}}]{Dasgupta:2018nvj}%
  \BibitemOpen
  \bibfield  {author} {\bibinfo {author} {\bibfnamefont {M.}~\bibnamefont
  {Dasgupta}}, \bibinfo {author} {\bibfnamefont {F.~A.}\ \bibnamefont
  {Dreyer}}, \bibinfo {author} {\bibfnamefont {K.}~\bibnamefont {Hamilton}},
  \bibinfo {author} {\bibfnamefont {P.~F.}\ \bibnamefont {Monni}}, \ and\
  \bibinfo {author} {\bibfnamefont {G.~P.}\ \bibnamefont {Salam}},\ }\href
  {\doibase 10.1007/JHEP09(2018)033} {\bibfield  {journal} {\bibinfo  {journal}
  {JHEP}\ }\textbf {\bibinfo {volume} {09}},\ \bibinfo {pages} {033} (\bibinfo
  {year} {2018})},\ \bibinfo {note} {[Erratum: JHEP 03, 083 (2020)]},\ \Eprint
  {http://arxiv.org/abs/1805.09327} {arXiv:1805.09327 [hep-ph]} \BibitemShut
  {NoStop}%
\bibitem [{\citenamefont {Forshaw}\ \emph {et~al.}(2019)\citenamefont
  {Forshaw}, \citenamefont {Holguin},\ and\ \citenamefont
  {Pl\"atzer}}]{Forshaw:2019ver}%
  \BibitemOpen
  \bibfield  {author} {\bibinfo {author} {\bibfnamefont {J.~R.}\ \bibnamefont
  {Forshaw}}, \bibinfo {author} {\bibfnamefont {J.}~\bibnamefont {Holguin}}, \
  and\ \bibinfo {author} {\bibfnamefont {S.}~\bibnamefont {Pl\"atzer}},\ }\href
  {\doibase 10.1007/JHEP08(2019)145} {\bibfield  {journal} {\bibinfo  {journal}
  {JHEP}\ }\textbf {\bibinfo {volume} {08}},\ \bibinfo {pages} {145} (\bibinfo
  {year} {2019})},\ \Eprint {http://arxiv.org/abs/1905.08686} {arXiv:1905.08686
  [hep-ph]} \BibitemShut {NoStop}%
\bibitem [{\citenamefont {Nagy}\ and\ \citenamefont
  {Soper}(2019)}]{Nagy:2019rwb}%
  \BibitemOpen
  \bibfield  {author} {\bibinfo {author} {\bibfnamefont {Z.}~\bibnamefont
  {Nagy}}\ and\ \bibinfo {author} {\bibfnamefont {D.~E.}\ \bibnamefont
  {Soper}},\ }\href {\doibase 10.1103/PhysRevD.100.074005} {\bibfield
  {journal} {\bibinfo  {journal} {Phys. Rev. D}\ }\textbf {\bibinfo {volume}
  {100}},\ \bibinfo {pages} {074005} (\bibinfo {year} {2019})},\ \Eprint
  {http://arxiv.org/abs/1908.11420} {arXiv:1908.11420 [hep-ph]} \BibitemShut
  {NoStop}%
\bibitem [{\citenamefont {H\"oche}\ and\ \citenamefont
  {Reichelt}(2020)}]{Hoeche:2020nsx}%
  \BibitemOpen
  \bibfield  {author} {\bibinfo {author} {\bibfnamefont {S.}~\bibnamefont
  {H\"oche}}\ and\ \bibinfo {author} {\bibfnamefont {D.}~\bibnamefont
  {Reichelt}},\ }\href@noop {} {\  (\bibinfo {year} {2020})},\ \Eprint
  {http://arxiv.org/abs/2001.11492} {arXiv:2001.11492 [hep-ph]} \BibitemShut
  {NoStop}%
\bibitem [{\citenamefont {Dasgupta}\ \emph {et~al.}(2020)\citenamefont
  {Dasgupta}, \citenamefont {Dreyer}, \citenamefont {Hamilton}, \citenamefont
  {Monni}, \citenamefont {Salam},\ and\ \citenamefont
  {Soyez}}]{Dasgupta:2020fwr}%
  \BibitemOpen
  \bibfield  {author} {\bibinfo {author} {\bibfnamefont {M.}~\bibnamefont
  {Dasgupta}}, \bibinfo {author} {\bibfnamefont {F.~A.}\ \bibnamefont
  {Dreyer}}, \bibinfo {author} {\bibfnamefont {K.}~\bibnamefont {Hamilton}},
  \bibinfo {author} {\bibfnamefont {P.~F.}\ \bibnamefont {Monni}}, \bibinfo
  {author} {\bibfnamefont {G.~P.}\ \bibnamefont {Salam}}, \ and\ \bibinfo
  {author} {\bibfnamefont {G.}~\bibnamefont {Soyez}},\ }\href {\doibase
  10.1103/PhysRevLett.125.052002} {\bibfield  {journal} {\bibinfo  {journal}
  {Phys. Rev. Lett.}\ }\textbf {\bibinfo {volume} {125}},\ \bibinfo {pages}
  {052002} (\bibinfo {year} {2020})},\ \Eprint
  {http://arxiv.org/abs/2002.11114} {arXiv:2002.11114 [hep-ph]} \BibitemShut
  {NoStop}%
\bibitem [{\citenamefont {Balsiger}\ \emph {et~al.}(2020)\citenamefont
  {Balsiger}, \citenamefont {Becher},\ and\ \citenamefont
  {Ferroglia}}]{Balsiger:2020ogy}%
  \BibitemOpen
  \bibfield  {author} {\bibinfo {author} {\bibfnamefont {M.}~\bibnamefont
  {Balsiger}}, \bibinfo {author} {\bibfnamefont {T.}~\bibnamefont {Becher}}, \
  and\ \bibinfo {author} {\bibfnamefont {A.}~\bibnamefont {Ferroglia}},\ }\href
  {\doibase 10.1007/JHEP09(2020)029} {\bibfield  {journal} {\bibinfo  {journal}
  {JHEP}\ }\textbf {\bibinfo {volume} {09}},\ \bibinfo {pages} {029} (\bibinfo
  {year} {2020})},\ \Eprint {http://arxiv.org/abs/2006.00014} {arXiv:2006.00014
  [hep-ph]} \BibitemShut {NoStop}%
\bibitem [{\citenamefont {Angelis}\ \emph {et~al.}(2020)\citenamefont
  {Angelis}, \citenamefont {Forshaw},\ and\ \citenamefont
  {Pl\"atzer}}]{DeAngelis:2020rvq}%
  \BibitemOpen
  \bibfield  {author} {\bibinfo {author} {\bibfnamefont {M.~D.}\ \bibnamefont
  {Angelis}}, \bibinfo {author} {\bibfnamefont {J.~R.}\ \bibnamefont
  {Forshaw}}, \ and\ \bibinfo {author} {\bibfnamefont {S.}~\bibnamefont
  {Pl\"atzer}},\ }\href@noop {} {\  (\bibinfo {year} {2020})},\ \Eprint
  {http://arxiv.org/abs/2007.09648} {arXiv:2007.09648 [hep-ph]} \BibitemShut
  {NoStop}%
\bibitem [{\citenamefont {Hamilton}\ \emph {et~al.}(2020)\citenamefont
  {Hamilton}, \citenamefont {Medves}, \citenamefont {Salam}, \citenamefont
  {Scyboz},\ and\ \citenamefont {Soyez}}]{1831912}%
  \BibitemOpen
  \bibfield  {author} {\bibinfo {author} {\bibfnamefont {K.}~\bibnamefont
  {Hamilton}}, \bibinfo {author} {\bibfnamefont {R.}~\bibnamefont {Medves}},
  \bibinfo {author} {\bibfnamefont {G.~P.}\ \bibnamefont {Salam}}, \bibinfo
  {author} {\bibfnamefont {L.}~\bibnamefont {Scyboz}}, \ and\ \bibinfo {author}
  {\bibfnamefont {G.}~\bibnamefont {Soyez}},\ }\href@noop {} {\  (\bibinfo
  {year} {2020})},\ \Eprint {http://arxiv.org/abs/2011.10054} {arXiv:2011.10054
  [hep-ph]} \BibitemShut {NoStop}%
\bibitem [{\citenamefont {H\"oche}(2015)}]{Hoche:2014rga}%
  \BibitemOpen
  \bibfield  {author} {\bibinfo {author} {\bibfnamefont {S.}~\bibnamefont
  {H\"oche}},\ }in\ \href {\doibase 10.1142/9789814678766_0005} {\emph
  {\bibinfo {booktitle} {{Theoretical Advanced Study Institute in Elementary
  Particle Physics}: {Journeys Through the Precision Frontier: Amplitudes for
  Colliders}}}}\ (\bibinfo {year} {2015})\ pp.\ \bibinfo {pages} {235--295},\
  \Eprint {http://arxiv.org/abs/1411.4085} {arXiv:1411.4085 [hep-ph]}
  \BibitemShut {NoStop}%
\bibitem [{\citenamefont {Dasgupta}\ and\ \citenamefont
  {Salam}(2001)}]{Dasgupta:2001sh}%
  \BibitemOpen
  \bibfield  {author} {\bibinfo {author} {\bibfnamefont {M.}~\bibnamefont
  {Dasgupta}}\ and\ \bibinfo {author} {\bibfnamefont {G.}~\bibnamefont
  {Salam}},\ }\href {\doibase 10.1016/S0370-2693(01)00725-0} {\bibfield
  {journal} {\bibinfo  {journal} {Phys. Lett. B}\ }\textbf {\bibinfo {volume}
  {512}},\ \bibinfo {pages} {323} (\bibinfo {year} {2001})},\ \Eprint
  {http://arxiv.org/abs/hep-ph/0104277} {arXiv:hep-ph/0104277} \BibitemShut
  {NoStop}%
\bibitem [{\citenamefont {Dasgupta}\ and\ \citenamefont
  {Salam}(2002)}]{Dasgupta:2002bw}%
  \BibitemOpen
  \bibfield  {author} {\bibinfo {author} {\bibfnamefont {M.}~\bibnamefont
  {Dasgupta}}\ and\ \bibinfo {author} {\bibfnamefont {G.~P.}\ \bibnamefont
  {Salam}},\ }\href {\doibase 10.1088/1126-6708/2002/03/017} {\bibfield
  {journal} {\bibinfo  {journal} {JHEP}\ }\textbf {\bibinfo {volume} {03}},\
  \bibinfo {pages} {017} (\bibinfo {year} {2002})},\ \Eprint
  {http://arxiv.org/abs/hep-ph/0203009} {arXiv:hep-ph/0203009} \BibitemShut
  {NoStop}%
\bibitem [{\citenamefont {Banfi}\ \emph {et~al.}(2002)\citenamefont {Banfi},
  \citenamefont {Marchesini},\ and\ \citenamefont {Smye}}]{Banfi:2002hw}%
  \BibitemOpen
  \bibfield  {author} {\bibinfo {author} {\bibfnamefont {A.}~\bibnamefont
  {Banfi}}, \bibinfo {author} {\bibfnamefont {G.}~\bibnamefont {Marchesini}}, \
  and\ \bibinfo {author} {\bibfnamefont {G.}~\bibnamefont {Smye}},\ }\href
  {\doibase 10.1088/1126-6708/2002/08/006} {\bibfield  {journal} {\bibinfo
  {journal} {JHEP}\ }\textbf {\bibinfo {volume} {08}},\ \bibinfo {pages} {006}
  (\bibinfo {year} {2002})},\ \Eprint {http://arxiv.org/abs/hep-ph/0206076}
  {arXiv:hep-ph/0206076} \BibitemShut {NoStop}%
\bibitem [{\citenamefont {Hatta}\ and\ \citenamefont
  {Ueda}(2013)}]{Hatta:2013iba}%
  \BibitemOpen
  \bibfield  {author} {\bibinfo {author} {\bibfnamefont {Y.}~\bibnamefont
  {Hatta}}\ and\ \bibinfo {author} {\bibfnamefont {T.}~\bibnamefont {Ueda}},\
  }\href {\doibase 10.1016/j.nuclphysb.2013.06.021} {\bibfield  {journal}
  {\bibinfo  {journal} {Nucl. Phys. B}\ }\textbf {\bibinfo {volume} {874}},\
  \bibinfo {pages} {808} (\bibinfo {year} {2013})},\ \Eprint
  {http://arxiv.org/abs/1304.6930} {arXiv:1304.6930 [hep-ph]} \BibitemShut
  {NoStop}%
\bibitem [{\citenamefont {Weigert}(2004)}]{Weigert:2003mm}%
  \BibitemOpen
  \bibfield  {author} {\bibinfo {author} {\bibfnamefont {H.}~\bibnamefont
  {Weigert}},\ }\href {\doibase 10.1016/j.nuclphysb.2004.03.002} {\bibfield
  {journal} {\bibinfo  {journal} {Nucl. Phys. B}\ }\textbf {\bibinfo {volume}
  {685}},\ \bibinfo {pages} {321} (\bibinfo {year} {2004})},\ \Eprint
  {http://arxiv.org/abs/hep-ph/0312050} {arXiv:hep-ph/0312050} \BibitemShut
  {NoStop}%
\bibitem [{\citenamefont {Hatta}(2008)}]{Hatta:2008st}%
  \BibitemOpen
  \bibfield  {author} {\bibinfo {author} {\bibfnamefont {Y.}~\bibnamefont
  {Hatta}},\ }\href {\doibase 10.1088/1126-6708/2008/11/057} {\bibfield
  {journal} {\bibinfo  {journal} {JHEP}\ }\textbf {\bibinfo {volume} {11}},\
  \bibinfo {pages} {057} (\bibinfo {year} {2008})},\ \Eprint
  {http://arxiv.org/abs/0810.0889} {arXiv:0810.0889 [hep-ph]} \BibitemShut
  {NoStop}%
\bibitem [{\citenamefont {Caron-Huot}(2018)}]{Caron-Huot:2015bja}%
  \BibitemOpen
  \bibfield  {author} {\bibinfo {author} {\bibfnamefont {S.}~\bibnamefont
  {Caron-Huot}},\ }\href {\doibase 10.1007/JHEP03(2018)036} {\bibfield
  {journal} {\bibinfo  {journal} {JHEP}\ }\textbf {\bibinfo {volume} {03}},\
  \bibinfo {pages} {036} (\bibinfo {year} {2018})},\ \Eprint
  {http://arxiv.org/abs/1501.03754} {arXiv:1501.03754 [hep-ph]} \BibitemShut
  {NoStop}%
\bibitem [{\citenamefont {Neill}\ and\ \citenamefont
  {Ringer}(2020)}]{Neill:2020bwv}%
  \BibitemOpen
  \bibfield  {author} {\bibinfo {author} {\bibfnamefont {D.}~\bibnamefont
  {Neill}}\ and\ \bibinfo {author} {\bibfnamefont {F.}~\bibnamefont {Ringer}},\
  }\href {\doibase 10.1007/JHEP06(2020)086} {\bibfield  {journal} {\bibinfo
  {journal} {JHEP}\ }\textbf {\bibinfo {volume} {06}},\ \bibinfo {pages} {086}
  (\bibinfo {year} {2020})},\ \Eprint {http://arxiv.org/abs/2003.02275}
  {arXiv:2003.02275 [hep-ph]} \BibitemShut {NoStop}%
\bibitem [{\citenamefont {Blaizot}\ \emph {et~al.}(2003)\citenamefont
  {Blaizot}, \citenamefont {Iancu},\ and\ \citenamefont
  {Weigert}}]{Blaizot:2002np}%
  \BibitemOpen
  \bibfield  {author} {\bibinfo {author} {\bibfnamefont {J.-P.}\ \bibnamefont
  {Blaizot}}, \bibinfo {author} {\bibfnamefont {E.}~\bibnamefont {Iancu}}, \
  and\ \bibinfo {author} {\bibfnamefont {H.}~\bibnamefont {Weigert}},\ }\href
  {\doibase 10.1016/S0375-9474(02)01299-X} {\bibfield  {journal} {\bibinfo
  {journal} {Nucl. Phys. A}\ }\textbf {\bibinfo {volume} {713}},\ \bibinfo
  {pages} {441} (\bibinfo {year} {2003})},\ \Eprint
  {http://arxiv.org/abs/hep-ph/0206279} {arXiv:hep-ph/0206279} \BibitemShut
  {NoStop}%
\bibitem [{\citenamefont {Hagiwara}\ \emph {et~al.}(2016)\citenamefont
  {Hagiwara}, \citenamefont {Hatta},\ and\ \citenamefont
  {Ueda}}]{Hagiwara:2015bia}%
  \BibitemOpen
  \bibfield  {author} {\bibinfo {author} {\bibfnamefont {Y.}~\bibnamefont
  {Hagiwara}}, \bibinfo {author} {\bibfnamefont {Y.}~\bibnamefont {Hatta}}, \
  and\ \bibinfo {author} {\bibfnamefont {T.}~\bibnamefont {Ueda}},\ }\href
  {\doibase 10.1016/j.physletb.2016.03.028} {\bibfield  {journal} {\bibinfo
  {journal} {Phys. Lett. B}\ }\textbf {\bibinfo {volume} {756}},\ \bibinfo
  {pages} {254} (\bibinfo {year} {2016})},\ \Eprint
  {http://arxiv.org/abs/1507.07641} {arXiv:1507.07641 [hep-ph]} \BibitemShut
  {NoStop}%
\bibitem [{\citenamefont {Hatta}\ \emph {et~al.}(2013)\citenamefont {Hatta},
  \citenamefont {Marquet}, \citenamefont {Royon}, \citenamefont {Soyez},
  \citenamefont {Ueda},\ and\ \citenamefont {Werder}}]{Hatta:2013qj}%
  \BibitemOpen
  \bibfield  {author} {\bibinfo {author} {\bibfnamefont {Y.}~\bibnamefont
  {Hatta}}, \bibinfo {author} {\bibfnamefont {C.}~\bibnamefont {Marquet}},
  \bibinfo {author} {\bibfnamefont {C.}~\bibnamefont {Royon}}, \bibinfo
  {author} {\bibfnamefont {G.}~\bibnamefont {Soyez}}, \bibinfo {author}
  {\bibfnamefont {T.}~\bibnamefont {Ueda}}, \ and\ \bibinfo {author}
  {\bibfnamefont {D.}~\bibnamefont {Werder}},\ }\href {\doibase
  10.1103/PhysRevD.87.054016} {\bibfield  {journal} {\bibinfo  {journal} {Phys.
  Rev. D}\ }\textbf {\bibinfo {volume} {87}},\ \bibinfo {pages} {054016}
  (\bibinfo {year} {2013})},\ \Eprint {http://arxiv.org/abs/1301.1910}
  {arXiv:1301.1910 [hep-ph]} \BibitemShut {NoStop}%
\bibitem [{\citenamefont {Hatta}\ and\ \citenamefont
  {Ueda}(2009)}]{Hatta:2009nd}%
  \BibitemOpen
  \bibfield  {author} {\bibinfo {author} {\bibfnamefont {Y.}~\bibnamefont
  {Hatta}}\ and\ \bibinfo {author} {\bibfnamefont {T.}~\bibnamefont {Ueda}},\
  }\href {\doibase 10.1103/PhysRevD.80.074018} {\bibfield  {journal} {\bibinfo
  {journal} {Phys. Rev. D}\ }\textbf {\bibinfo {volume} {80}},\ \bibinfo
  {pages} {074018} (\bibinfo {year} {2009})},\ \Eprint
  {http://arxiv.org/abs/0909.0056} {arXiv:0909.0056 [hep-ph]} \BibitemShut
  {NoStop}%
\bibitem [{\citenamefont {Balitsky}(1996)}]{Balitsky:1995ub}%
  \BibitemOpen
  \bibfield  {author} {\bibinfo {author} {\bibfnamefont {I.}~\bibnamefont
  {Balitsky}},\ }\href {\doibase 10.1016/0550-3213(95)00638-9} {\bibfield
  {journal} {\bibinfo  {journal} {Nucl. Phys. B}\ }\textbf {\bibinfo {volume}
  {463}},\ \bibinfo {pages} {99} (\bibinfo {year} {1996})},\ \Eprint
  {http://arxiv.org/abs/hep-ph/9509348} {arXiv:hep-ph/9509348} \BibitemShut
  {NoStop}%
\bibitem [{\citenamefont {Kovchegov}\ \emph {et~al.}(2009)\citenamefont
  {Kovchegov}, \citenamefont {Kuokkanen}, \citenamefont {Rummukainen},\ and\
  \citenamefont {Weigert}}]{Kovchegov:2008mk}%
  \BibitemOpen
  \bibfield  {author} {\bibinfo {author} {\bibfnamefont {Y.~V.}\ \bibnamefont
  {Kovchegov}}, \bibinfo {author} {\bibfnamefont {J.}~\bibnamefont
  {Kuokkanen}}, \bibinfo {author} {\bibfnamefont {K.}~\bibnamefont
  {Rummukainen}}, \ and\ \bibinfo {author} {\bibfnamefont {H.}~\bibnamefont
  {Weigert}},\ }\href {\doibase 10.1016/j.nuclphysa.2009.03.006} {\bibfield
  {journal} {\bibinfo  {journal} {Nucl. Phys. A}\ }\textbf {\bibinfo {volume}
  {823}},\ \bibinfo {pages} {47} (\bibinfo {year} {2009})},\ \Eprint
  {http://arxiv.org/abs/0812.3238} {arXiv:0812.3238 [hep-ph]} \BibitemShut
  {NoStop}%
\bibitem [{\citenamefont {Dominguez}\ \emph
  {et~al.}(2011{\natexlab{a}})\citenamefont {Dominguez}, \citenamefont
  {Mueller}, \citenamefont {Munier},\ and\ \citenamefont
  {Xiao}}]{Dominguez:2011gc}%
  \BibitemOpen
  \bibfield  {author} {\bibinfo {author} {\bibfnamefont {F.}~\bibnamefont
  {Dominguez}}, \bibinfo {author} {\bibfnamefont {A.}~\bibnamefont {Mueller}},
  \bibinfo {author} {\bibfnamefont {S.}~\bibnamefont {Munier}}, \ and\ \bibinfo
  {author} {\bibfnamefont {B.-W.}\ \bibnamefont {Xiao}},\ }\href {\doibase
  10.1016/j.physletb.2011.09.104} {\bibfield  {journal} {\bibinfo  {journal}
  {Phys. Lett. B}\ }\textbf {\bibinfo {volume} {705}},\ \bibinfo {pages} {106}
  (\bibinfo {year} {2011}{\natexlab{a}})},\ \Eprint
  {http://arxiv.org/abs/1108.1752} {arXiv:1108.1752 [hep-ph]} \BibitemShut
  {NoStop}%
\bibitem [{\citenamefont {Dumitru}\ \emph {et~al.}(2011)\citenamefont
  {Dumitru}, \citenamefont {Jalilian-Marian}, \citenamefont {Lappi},
  \citenamefont {Schenke},\ and\ \citenamefont {Venugopalan}}]{Dumitru:2011vk}%
  \BibitemOpen
  \bibfield  {author} {\bibinfo {author} {\bibfnamefont {A.}~\bibnamefont
  {Dumitru}}, \bibinfo {author} {\bibfnamefont {J.}~\bibnamefont
  {Jalilian-Marian}}, \bibinfo {author} {\bibfnamefont {T.}~\bibnamefont
  {Lappi}}, \bibinfo {author} {\bibfnamefont {B.}~\bibnamefont {Schenke}}, \
  and\ \bibinfo {author} {\bibfnamefont {R.}~\bibnamefont {Venugopalan}},\
  }\href {\doibase 10.1016/j.physletb.2011.11.002} {\bibfield  {journal}
  {\bibinfo  {journal} {Phys. Lett. B}\ }\textbf {\bibinfo {volume} {706}},\
  \bibinfo {pages} {219} (\bibinfo {year} {2011})},\ \Eprint
  {http://arxiv.org/abs/1108.4764} {arXiv:1108.4764 [hep-ph]} \BibitemShut
  {NoStop}%
\bibitem [{\citenamefont {Marquet}\ \emph {et~al.}(2016)\citenamefont
  {Marquet}, \citenamefont {Petreska},\ and\ \citenamefont
  {Roiesnel}}]{Marquet:2016cgx}%
  \BibitemOpen
  \bibfield  {author} {\bibinfo {author} {\bibfnamefont {C.}~\bibnamefont
  {Marquet}}, \bibinfo {author} {\bibfnamefont {E.}~\bibnamefont {Petreska}}, \
  and\ \bibinfo {author} {\bibfnamefont {C.}~\bibnamefont {Roiesnel}},\ }\href
  {\doibase 10.1007/JHEP10(2016)065} {\bibfield  {journal} {\bibinfo  {journal}
  {JHEP}\ }\textbf {\bibinfo {volume} {10}},\ \bibinfo {pages} {065} (\bibinfo
  {year} {2016})},\ \Eprint {http://arxiv.org/abs/1608.02577} {arXiv:1608.02577
  [hep-ph]} \BibitemShut {NoStop}%
\bibitem [{\citenamefont {Forshaw}\ \emph {et~al.}(2006)\citenamefont
  {Forshaw}, \citenamefont {Kyrieleis},\ and\ \citenamefont
  {Seymour}}]{Forshaw:2006fk}%
  \BibitemOpen
  \bibfield  {author} {\bibinfo {author} {\bibfnamefont {J.~R.}\ \bibnamefont
  {Forshaw}}, \bibinfo {author} {\bibfnamefont {A.}~\bibnamefont {Kyrieleis}},
  \ and\ \bibinfo {author} {\bibfnamefont {M.}~\bibnamefont {Seymour}},\ }\href
  {\doibase 10.1088/1126-6708/2006/08/059} {\bibfield  {journal} {\bibinfo
  {journal} {JHEP}\ }\textbf {\bibinfo {volume} {08}},\ \bibinfo {pages} {059}
  (\bibinfo {year} {2006})},\ \Eprint {http://arxiv.org/abs/hep-ph/0604094}
  {arXiv:hep-ph/0604094} \BibitemShut {NoStop}%
\bibitem [{\citenamefont {Forshaw}\ \emph {et~al.}(2008)\citenamefont
  {Forshaw}, \citenamefont {Kyrieleis},\ and\ \citenamefont
  {Seymour}}]{Forshaw:2008cq}%
  \BibitemOpen
  \bibfield  {author} {\bibinfo {author} {\bibfnamefont {J.}~\bibnamefont
  {Forshaw}}, \bibinfo {author} {\bibfnamefont {A.}~\bibnamefont {Kyrieleis}},
  \ and\ \bibinfo {author} {\bibfnamefont {M.}~\bibnamefont {Seymour}},\ }\href
  {\doibase 10.1088/1126-6708/2008/09/128} {\bibfield  {journal} {\bibinfo
  {journal} {JHEP}\ }\textbf {\bibinfo {volume} {09}},\ \bibinfo {pages} {128}
  (\bibinfo {year} {2008})},\ \Eprint {http://arxiv.org/abs/0808.1269}
  {arXiv:0808.1269 [hep-ph]} \BibitemShut {NoStop}%
\bibitem [{\citenamefont {Mueller}\ and\ \citenamefont
  {Patel}(1994)}]{Mueller:1994jq}%
  \BibitemOpen
  \bibfield  {author} {\bibinfo {author} {\bibfnamefont {A.~H.}\ \bibnamefont
  {Mueller}}\ and\ \bibinfo {author} {\bibfnamefont {B.}~\bibnamefont
  {Patel}},\ }\href {\doibase 10.1016/0550-3213(94)90284-4} {\bibfield
  {journal} {\bibinfo  {journal} {Nucl. Phys. B}\ }\textbf {\bibinfo {volume}
  {425}},\ \bibinfo {pages} {471} (\bibinfo {year} {1994})},\ \Eprint
  {http://arxiv.org/abs/hep-ph/9403256} {arXiv:hep-ph/9403256} \BibitemShut
  {NoStop}%
\bibitem [{\citenamefont {Hatta}\ and\ \citenamefont
  {Mueller}(2007)}]{Hatta:2007fg}%
  \BibitemOpen
  \bibfield  {author} {\bibinfo {author} {\bibfnamefont {Y.}~\bibnamefont
  {Hatta}}\ and\ \bibinfo {author} {\bibfnamefont {A.}~\bibnamefont
  {Mueller}},\ }\href {\doibase 10.1016/j.nuclphysa.2007.03.003} {\bibfield
  {journal} {\bibinfo  {journal} {Nucl. Phys. A}\ }\textbf {\bibinfo {volume}
  {789}},\ \bibinfo {pages} {285} (\bibinfo {year} {2007})},\ \Eprint
  {http://arxiv.org/abs/hep-ph/0702023} {arXiv:hep-ph/0702023} \BibitemShut
  {NoStop}%
\bibitem [{\citenamefont {Avsar}\ and\ \citenamefont
  {Hatta}(2008)}]{Avsar:2008ph}%
  \BibitemOpen
  \bibfield  {author} {\bibinfo {author} {\bibfnamefont {E.}~\bibnamefont
  {Avsar}}\ and\ \bibinfo {author} {\bibfnamefont {Y.}~\bibnamefont {Hatta}},\
  }\href {\doibase 10.1088/1126-6708/2008/09/102} {\bibfield  {journal}
  {\bibinfo  {journal} {JHEP}\ }\textbf {\bibinfo {volume} {09}},\ \bibinfo
  {pages} {102} (\bibinfo {year} {2008})},\ \Eprint
  {http://arxiv.org/abs/0805.0710} {arXiv:0805.0710 [hep-ph]} \BibitemShut
  {NoStop}%
\bibitem [{\citenamefont {Avsar}\ \emph {et~al.}(2009)\citenamefont {Avsar},
  \citenamefont {Hatta},\ and\ \citenamefont {Matsuo}}]{Avsar:2009yb}%
  \BibitemOpen
  \bibfield  {author} {\bibinfo {author} {\bibfnamefont {E.}~\bibnamefont
  {Avsar}}, \bibinfo {author} {\bibfnamefont {Y.}~\bibnamefont {Hatta}}, \ and\
  \bibinfo {author} {\bibfnamefont {T.}~\bibnamefont {Matsuo}},\ }\href
  {\doibase 10.1088/1126-6708/2009/06/011} {\bibfield  {journal} {\bibinfo
  {journal} {JHEP}\ }\textbf {\bibinfo {volume} {06}},\ \bibinfo {pages} {011}
  (\bibinfo {year} {2009})},\ \Eprint {http://arxiv.org/abs/0903.4285}
  {arXiv:0903.4285 [hep-ph]} \BibitemShut {NoStop}%
\bibitem [{\citenamefont {Kidonakis}\ \emph {et~al.}(1998)\citenamefont
  {Kidonakis}, \citenamefont {Oderda},\ and\ \citenamefont
  {Sterman}}]{Kidonakis:1998nf}%
  \BibitemOpen
  \bibfield  {author} {\bibinfo {author} {\bibfnamefont {N.}~\bibnamefont
  {Kidonakis}}, \bibinfo {author} {\bibfnamefont {G.}~\bibnamefont {Oderda}}, \
  and\ \bibinfo {author} {\bibfnamefont {G.~F.}\ \bibnamefont {Sterman}},\
  }\href {\doibase 10.1016/S0550-3213(98)00441-6} {\bibfield  {journal}
  {\bibinfo  {journal} {Nucl. Phys. B}\ }\textbf {\bibinfo {volume} {531}},\
  \bibinfo {pages} {365} (\bibinfo {year} {1998})},\ \Eprint
  {http://arxiv.org/abs/hep-ph/9803241} {arXiv:hep-ph/9803241} \BibitemShut
  {NoStop}%
\bibitem [{\citenamefont {Forshaw}\ and\ \citenamefont
  {Sjodahl}(2007)}]{Forshaw:2007vb}%
  \BibitemOpen
  \bibfield  {author} {\bibinfo {author} {\bibfnamefont {J.~R.}\ \bibnamefont
  {Forshaw}}\ and\ \bibinfo {author} {\bibfnamefont {M.}~\bibnamefont
  {Sjodahl}},\ }\href {\doibase 10.1088/1126-6708/2007/09/119} {\bibfield
  {journal} {\bibinfo  {journal} {JHEP}\ }\textbf {\bibinfo {volume} {09}},\
  \bibinfo {pages} {119} (\bibinfo {year} {2007})},\ \Eprint
  {http://arxiv.org/abs/0705.1504} {arXiv:0705.1504 [hep-ph]} \BibitemShut
  {NoStop}%
\bibitem [{\citenamefont {Blaizot}\ \emph {et~al.}(2004)\citenamefont
  {Blaizot}, \citenamefont {Gelis},\ and\ \citenamefont
  {Venugopalan}}]{Blaizot:2004wv}%
  \BibitemOpen
  \bibfield  {author} {\bibinfo {author} {\bibfnamefont {J.~P.}\ \bibnamefont
  {Blaizot}}, \bibinfo {author} {\bibfnamefont {F.}~\bibnamefont {Gelis}}, \
  and\ \bibinfo {author} {\bibfnamefont {R.}~\bibnamefont {Venugopalan}},\
  }\href {\doibase 10.1016/j.nuclphysa.2004.07.006} {\bibfield  {journal}
  {\bibinfo  {journal} {Nucl. Phys. A}\ }\textbf {\bibinfo {volume} {743}},\
  \bibinfo {pages} {57} (\bibinfo {year} {2004})},\ \Eprint
  {http://arxiv.org/abs/hep-ph/0402257} {arXiv:hep-ph/0402257} \BibitemShut
  {NoStop}%
\bibitem [{\citenamefont {Dominguez}\ \emph
  {et~al.}(2011{\natexlab{b}})\citenamefont {Dominguez}, \citenamefont
  {Marquet}, \citenamefont {Xiao},\ and\ \citenamefont
  {Yuan}}]{Dominguez:2011wm}%
  \BibitemOpen
  \bibfield  {author} {\bibinfo {author} {\bibfnamefont {F.}~\bibnamefont
  {Dominguez}}, \bibinfo {author} {\bibfnamefont {C.}~\bibnamefont {Marquet}},
  \bibinfo {author} {\bibfnamefont {B.-W.}\ \bibnamefont {Xiao}}, \ and\
  \bibinfo {author} {\bibfnamefont {F.}~\bibnamefont {Yuan}},\ }\href {\doibase
  10.1103/PhysRevD.83.105005} {\bibfield  {journal} {\bibinfo  {journal} {Phys.
  Rev. D}\ }\textbf {\bibinfo {volume} {83}},\ \bibinfo {pages} {105005}
  (\bibinfo {year} {2011}{\natexlab{b}})},\ \Eprint
  {http://arxiv.org/abs/1101.0715} {arXiv:1101.0715 [hep-ph]} \BibitemShut
  {NoStop}%
\bibitem [{\citenamefont {Shi}\ \emph {et~al.}(2017)\citenamefont {Shi},
  \citenamefont {Zhang},\ and\ \citenamefont {Wang}}]{Shi:2017gcq}%
  \BibitemOpen
  \bibfield  {author} {\bibinfo {author} {\bibfnamefont {Y.}~\bibnamefont
  {Shi}}, \bibinfo {author} {\bibfnamefont {C.}~\bibnamefont {Zhang}}, \ and\
  \bibinfo {author} {\bibfnamefont {E.}~\bibnamefont {Wang}},\ }\href {\doibase
  10.1103/PhysRevD.95.116014} {\bibfield  {journal} {\bibinfo  {journal} {Phys.
  Rev. D}\ }\textbf {\bibinfo {volume} {95}},\ \bibinfo {pages} {116014}
  (\bibinfo {year} {2017})},\ \Eprint {http://arxiv.org/abs/1704.00266}
  {arXiv:1704.00266 [hep-th]} \BibitemShut {NoStop}%
\end{thebibliography}%

\end{document}